\documentclass[twocolumn,preprintnumbers,superscriptaddress,amsmath,amssymb,prb]{revtex4-2}

\usepackage{graphicx}
\usepackage{wasysym}
\usepackage{tikz}
\usepackage{amsmath}
\usepackage[caption=false]{subfig}
\usepackage{bm}
\usepackage[utf8]{inputenc}
\usepackage{comment}
\usepackage[unicode=true, bookmarks=false, breaklinks=false, pdfborder={0 0 1},colorlinks=false]{hyperref}
\usepackage{breakurl}
\usepackage{url}
\usepackage{hyperref}
\usepackage{natbib}
\usepackage{multirow}
\usepackage{float}
\usepackage{appendix}
\usepackage{wrapfig}
\usepackage{xcolor}
\usepackage[most]{tcolorbox}

\begin{document}

\title{
When the magnetic force theorem fails: dressed exchange in the induced-moment magnets FePt, FePd, CoPt, and CoPd
}

\newcommand{\Uppsala}{Department of Physics and Astronomy, Uppsala University, Box 516, SE-75120 Uppsala, Sweden}
\newcommand{\KTH}{Department of Applied Physics, School of Engineering Sciences, KTH Royal Institute of Technology, AlbaNova University Center, SE-10691 Stockholm, Sweden}
\newcommand{\WISEUppsala}{Wallenberg Initiative Materials Science for Sustainability, Uppsala University,
75121 Uppsala, Sweden}
\newcommand{\WISEKTH}{Wallenberg Initiative Materials Science for Sustainability (WISE), KTH Royal Institute of Technology, SE-10044 Stockholm, Sweden}
\newcommand{\SeRC}{SeRC (Swedish e-Science Research Center), KTH Royal Institute of Technology, SE-10044 Stockholm, Sweden}
\newcommand{\Nijmegen}{Institute for Molecules and Materials, Radboud University, Heijendaalseweg 135, 6525 AJ Nijmegen, The Netherlands}
\newcommand{\Skovde}{The School of Engineering Science, University of Skövde, SE-541 28 Skövde, Sweden}
\newcommand{\Austin} {Department of Physics, University of Texas at Austin, Austin, TX, USA}

\author{Vladislav Borisov}
\affiliation{\Uppsala}
\affiliation{\WISEUppsala}
\email[Corresponding author:\ ]{vladislav.borisov@physics.uu.se}

\author{Maryna Pankratova}
\affiliation{\Uppsala}
\affiliation{\WISEUppsala}
\affiliation{\Skovde}

\author{Rohit Pathak}
\affiliation{\Uppsala}

\author{Pamela Costa Carvalho}
\affiliation{\Uppsala}

\author{Erna K. Delczeg-Czirjak}
\affiliation{\Uppsala}
\affiliation{\WISEUppsala}

\author{Sagar Sarkar}
\affiliation{\Uppsala}

\author{Anders Bergman}
\affiliation{\Uppsala}
\affiliation{\WISEUppsala}

\author{Qichen Xu}
    \affiliation{\KTH}
    \affiliation{\SeRC}

\author{Nezhat Pournaghavi}
    \affiliation{\Austin}
    \affiliation{\Uppsala}
    \affiliation{\KTH}

\author{Manuel Pereiro}
\affiliation{\Uppsala}    

\author{Hermann Dürr}
    \affiliation{\Uppsala}

\author{Mikhail I. Katsnelson}
\affiliation{\WISEUppsala}
\affiliation{\Nijmegen}
    
\author{Olle Eriksson}
\affiliation{\Uppsala}
\affiliation{\WISEUppsala}
    
\author{Anna Delin}
    \affiliation{\KTH}
    \affiliation{\SeRC}
    \affiliation{\WISEKTH}
    \email[Corresponding author:\ ]{annadel@kth.se}
    
\date{\today}

\begin{abstract}
Effective spin Hamiltonians for itinerant magnets are commonly
parameterized by magnetic-force-theorem (MFT) exchange constants,
which are subsequently used to compute the behavior of complex and
delicate spin textures, for example skyrmions. We show that for
systems containing atoms with large spin
susceptibility, such as Pd or Pt, this approach misses important parts of the
energetics. For FePt, it even predicts the wrong ground state.
Benchmarking against fully self-consistent spin-spiral
density-functional calculations --- in which the electronic
structure, including the induced Pd/Pt moments, relaxes for every
spin configuration --- we quantify the missing energetics for the
ordered L1$_0$ alloys FePt, FePd, CoPt, and CoPd. By comparing a
family of models that differ only in their treatment of the induced
Pt moment, including the configuration-dependent models of Mryasov
and Polesya, we identify the origin: the energetics of the induced
moments are governed by the exchange-enhanced (dressed)
susceptibility, of which the frozen-potential MFT captures only the
bare part --- about one third of the Pt-mediated interaction in
FePt. We derive a general bound showing that no on-site (intra-atomic)
refinement can supply the missing strength. The dressing decides the
magnetic ground state in FePt, where the interlayer ordering rests
entirely on the Pt-mediated interactions, and shifts the Curie
temperatures of the other alloys by several hundred kelvin. Computed
ordering temperatures, magnon spectra and spin wave stiffness for all four compounds are
provided.
\end{abstract}

\maketitle

\section{Introduction}
\begin{figure}[t]
\includegraphics[width=\columnwidth]{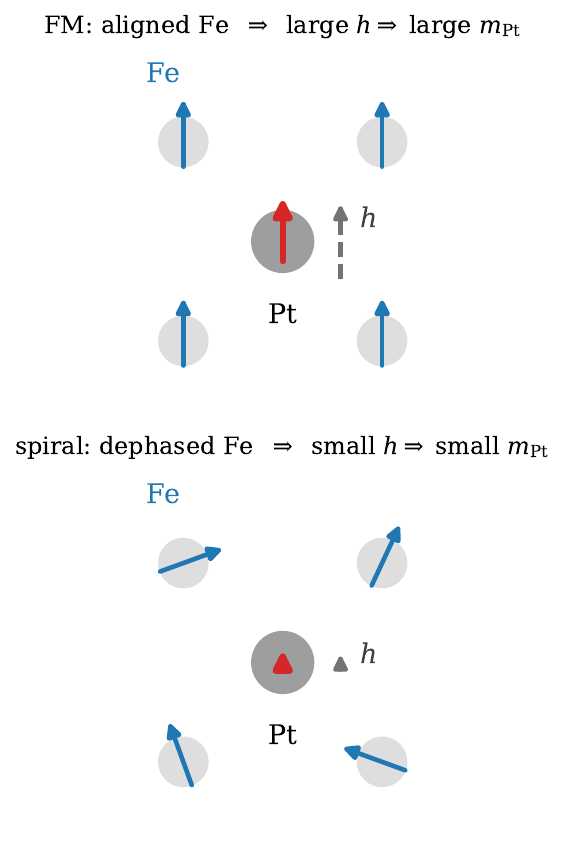}
  \caption{
  Schematic of the induced-Pt mechanism. The Pt moment (red arrow) is induced by the exchange field $h$ (grey arrow) of the surrounding Fe spins (blue arrows). Top: in the ferromagnetic (FM) state the aligned Fe produce a large field and hence a large induced Pt moment. Bottom: along a spin spiral (SS) the Fe moments dephase, the net field shrinks, and the Pt moment collapses.}
  \label{fig:Pt_moment_schematic}
\end{figure}
Effective spin Hamiltonians provide the standard description of magnetic materials by replacing the electronic degrees of freedom with interactions between localized magnetic moments. For many systems this mapping is highly successful, but it is not unique. The effective interactions depend on how the electronic response is incorporated. In elemental itinerant ferromagnets such as Fe and Ni this distinction is largely quantitative, leading to different parameter sets that are optimal for different observables, such as magnon spectra or thermodynamic properties \cite{Liechtenstein1984,LiechtensteinCurie1985,LKAG1987,Bruno2003,Katsnelson2004,singer2011,szilva2013,Cardias2022Comment,Daglum2025,Rezaei2025}. In the present work we show that for the L1$_0$ FePt system the distinction becomes qualitative. The conventional magnetic-force-theorem (MFT) interactions fail to reproduce even the correct magnetic ground state, so that neither the magnon spectrum nor the thermodynamics can be built upon them.
(For other L1$_0$ compounds the effects are quantitative but significant, with predicted critical temperatures off by several hundred degrees.)

The origin of this failure lies in the very different magnetic character of the Fe and Pt atoms. Fe carries a large, robust local moment that is well described as a rigid spin. Pt, in contrast, lies just below the Stoner instability. It has no spontaneous moment of its own but develops a sizeable induced moment in response to the exchange field generated by the surrounding Fe atoms, as illustrated in Fig.\,\ref{fig:Pt_moment_schematic}. The Pt moment is therefore not an independent magnetic degree of freedom, but a rapidly adjusting electronic response that follows the instantaneous configuration of the Fe moments.

This distinction can be described in terms of a hierarchy of magnetic susceptibilities. At the lowest level, the \emph{bare} susceptibility describes how the Pt spin density responds to the exchange field produced by the neighbouring Fe moments. The local Stoner interaction then amplifies this response by generating an additional exchange field on the same Pt atom. This local feedback does not remain local --- the induced polarization propagates through the crystal and generates further local exchange fields --- giving rise to an infinite hierarchy of propagation paths. Their complete resummation is described by the Dyson equation, yielding the \emph{dressed} susceptibility. 

Several effective spin models have been proposed to account for the induced Pt moment. In the model of Mryasov \textit{et al.},\cite{Mryasov2005} the Pt moments are described within linear response, which --- as we show below --- amounts to eliminating them analytically from an explicit spin Hamiltonian and yields a configuration-dependent Fe--Fe interaction, whereas Polesya \textit{et al.} \cite{Polesya2010} retain the Pt moments explicitly in an extended Heisenberg Hamiltonian. Both models incorporate the exchange-enhanced \emph{static} susceptibility (i.e., the response is evaluated after full electronic relaxation) of Pt. 
They therefore predict essentially the same induced Pt moments. Their differences arise instead from how this response is incorporated into the effective magnetic energy.

In this work we use fully self-consistent spin-spiral (SS) calculations based on the generalized Bloch theorem as the most accurate numerical reference for the configuration-dependent magnetic energetics of FePt. We compare these results with spin Hamiltonians constructed from magnetic-force-theorem (MFT) exchange parameters, as well as with the models of Mryasov and Polesya. By analyzing spin spirals along three complementary high-symmetry directions of the Brillouin zone (BZ), we isolate the role of the exchange-enhanced Pt response and show that it plays a central role in determining the magnetic order and therefore the qualitative magnetic behaviour. 
The resulting picture explains why conventional MFT exchange parameters fail for FePt and clarifies the respective roles of bare and dressed susceptibilities in constructing effective spin Hamiltonians for induced-moment systems.

Finally, we discuss the consequences of this picture for finite-temperature properties by calculating Curie temperatures, dynamical structure factors, and adiabatic magnon spectra for the complete family of ordered L1$_0$ alloys FePt, FePd, CoPt, and CoPd, comparing the results of the spin-spiral approach with the MFT approach.

\section{Results}

\renewcommand{\arraystretch}{1.5}
\setlength{\tabcolsep}{4pt}
\begin{table*}
  \caption{
  From left to right: Lattice parameters $a$ and $c$ used in the DFT calculations. Computed spin and orbital contributions ($m_s$ and $m_{\mathrm{orb}}$) to the magnetic moments at the Fe/Co and Pd/Pt sites (in $\mu_\mathrm{B}$), using RSPt. $K_\mathrm{U}$ is the uniaxial anisotropy constant per Fe/Co site (a negative value indicates easy-axis character), computed using RSPt. $\Delta E_\mathrm{magn}$ is the energy difference between the ferromagnetic (FM) state and antiferromagnetic one with antiparallel nearest-neighbor FM layers of Fe or Co, computed using RSPt. $m_s^\mathrm{total}$ is the total spin moment computed using the Elk software package, without spin-orbit coupling. This value is used together with the effective (Fe-Fe or Co-Co) interactions obtained from the spin-spiral computations. $T_\mathrm{C}^{\mathrm{MFT}}$ and
  $T_\mathrm{C}^{\mathrm{SS}}$ are computed Curie temperatures, estimated from Monte Carlo simulations (as implemented in the UppASD software package), using the MFT-based parameters and the spin-spiral-derived parameters, respectively. The full respective spin Hamiltonian was used in both cases. The right-most column shows experimental Curie temperatures taken from the literature. See discussion in the text for additional details.
  }
  \centering
  \vspace{5pt}
  \begin{tabular}{r|cc|cccc|l|c|c|lll}
    \hline
       & $a$ & $c$ & $m_s^\mathrm{Fe/Co}$ & $m_s^\mathrm{Pd/Pt}$ & $m_\mathrm{\mathrm{orb}}^\mathrm{Fe/Co}$ & $m_\mathrm{orb}^\mathrm{Pd/Pt}$ & $K_\mathrm{U}$ & $\Delta E_\mathrm{magn}$ & $m_s^\mathrm{total}$ & $T_\mathrm{C}^{\mathrm{MFT}}$ & $T_\mathrm{C}^{\mathrm{SS}}$ & $T_\mathrm{C}^{\mathrm{exp}}$ \\
    \hline
    units & Å & Å & $\mu_\mathrm{B}$ & $\mu_\mathrm{B}$ & $\mu_\mathrm{B}$ & $\mu_\mathrm{B}$ & meV/f.u. & meV/f.u. & $\mu_\mathrm{B}$ & K & K & K \\
    \hline
      FePt & 3.855 & 3.793 & 2.91 & 0.36 & 0.06 & 0.04 & $-2.49$ & $-1.9$ & 3.24 & AFM & 640 & $750$ \cite{LYUBINA2011291,klemmer1995magnetic,weller2000high} \\
      FePd & 3.859 & 3.716 & 3.00 & 0.37 & 0.07 & 0.03 & $-0.16$ & $-70.3$ & 3.29 & 525 & 906 & $734,\,760$ \cite{824418,kozubski2022atomic,wang2004effects,klemmer1995magnetic} \\      
      CoPt & 3.786 & 3.685 & 1.88 & 0.40 & 0.09 & 0.06 & $-0.95$ & $-36.1$ & 2.20 & 500 & 817 &  840 \cite{weller2000high,klemmer1995magnetic}  \\
      CoPd & 3.790 & 3.610 & 1.93 & 0.38 & 0.12 & 0.03 & $-0.016$ & $-48.0$ & 2.26  & 610 & 980 & $-$ \\
    \hline
  \end{tabular}
  \label{table:moments_TC}
\end{table*}

\begin{figure}
{\centering
\includegraphics[width=0.48\textwidth]{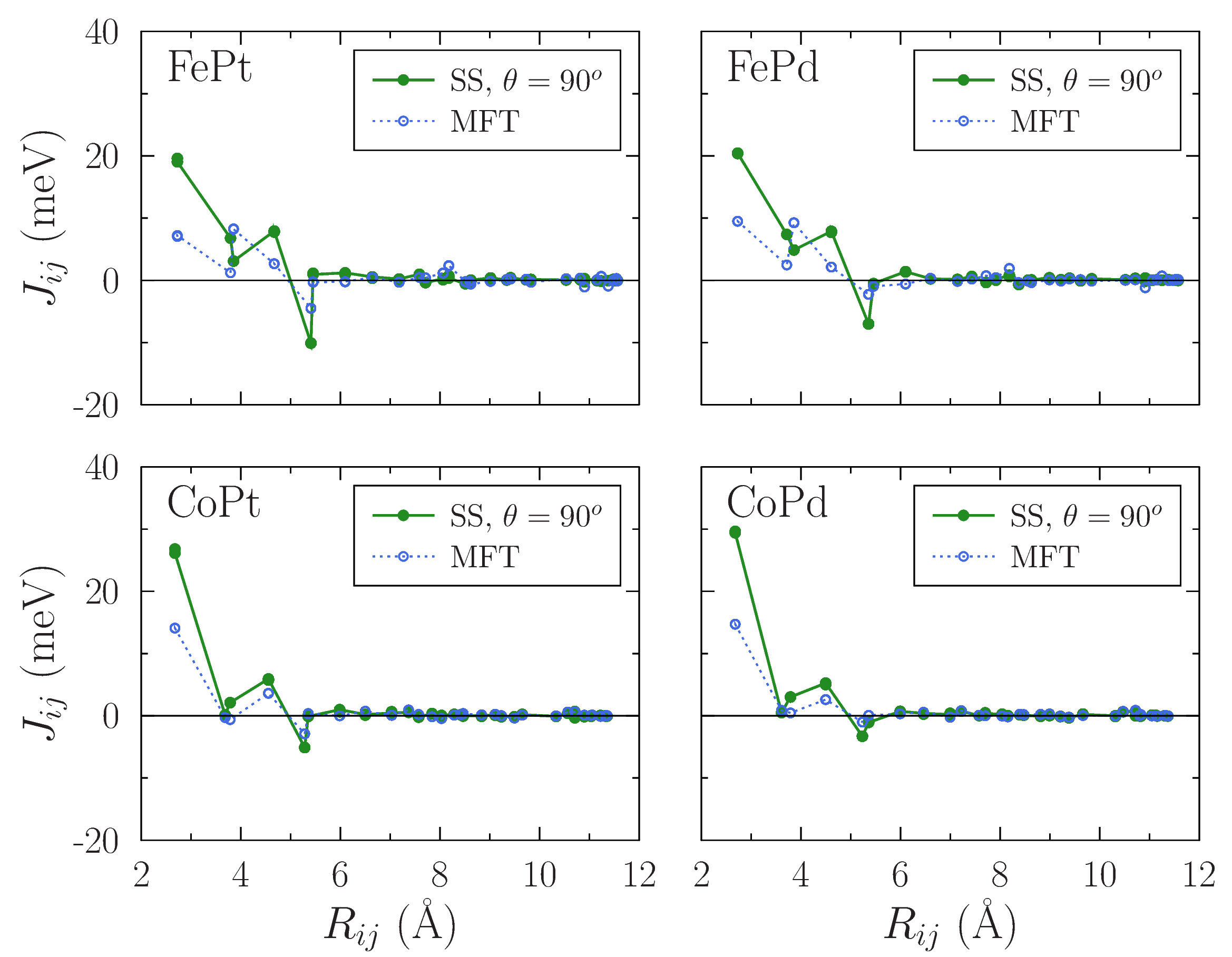}
}
\vspace{-20pt}
\caption{
Heisenberg interactions between $3d$ transition metal spins in the L1$_0$ structures calculated using the magnetic force theorem (MFT) (blue dashed lines, open symbols) or Fourier transform of the generalized Bloch spin spiral energies (green lines, filled symbols). Further interaction parameters (including Pt and Pd atoms) are summarized in Appendix~\ref{sec:further_interactions} and Fig.~\ref{f:all_Jijs}.
}
\label{fig:Jij_all}
\end{figure}

\begin{figure}[h]
{\centering
\includegraphics[width=0.4\textwidth]{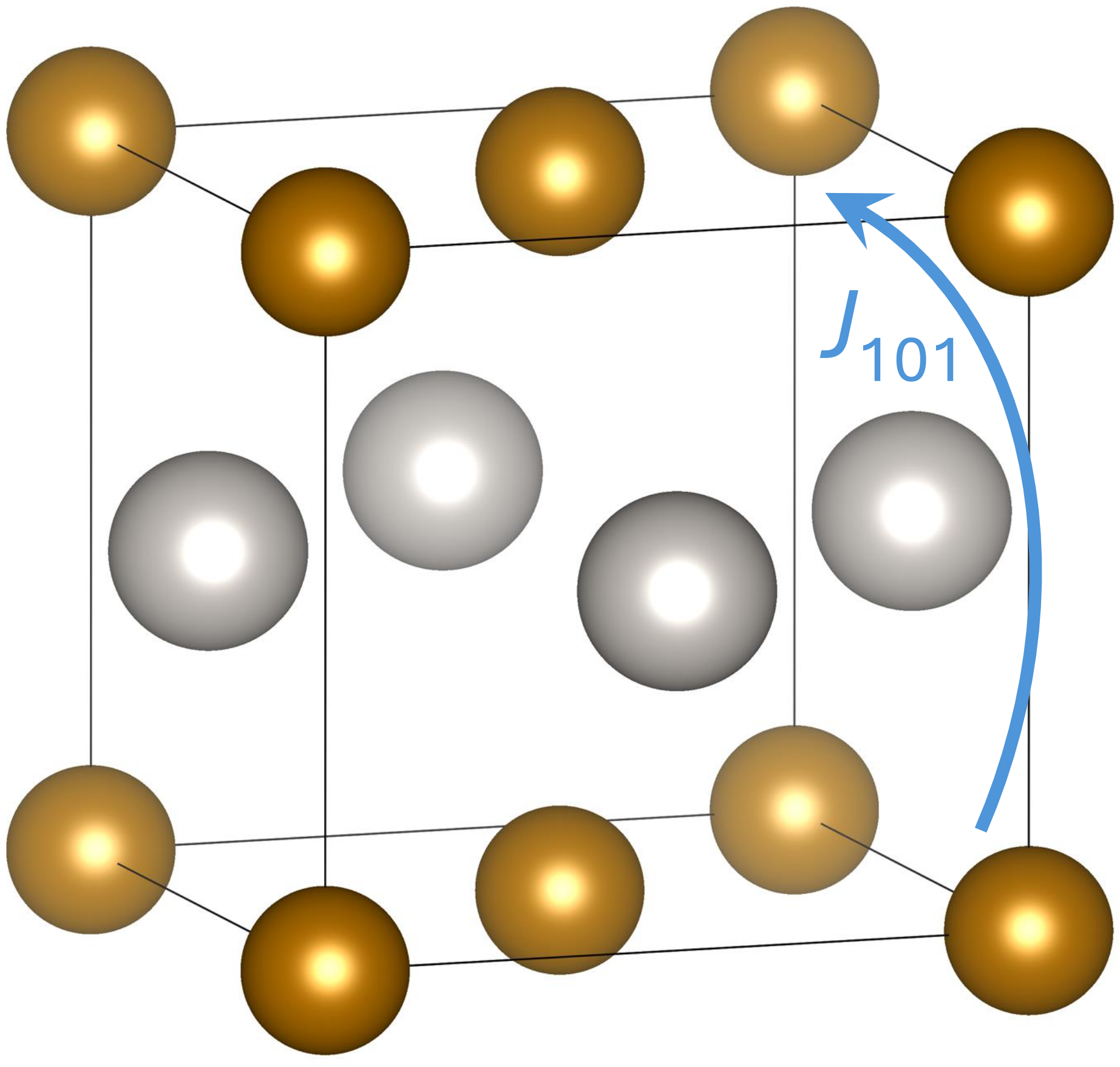}
}
\caption{
Crystal structure of the here studied L1$_0$ systems. The gray spheres are Pd/Pt atoms, whereas the brown spheres are Fe/Co atoms.
The blue arrow $J_{101}$ 
indicates the nearest antiferromagnetic Fe-Fe interaction, present in all four studied systems at $R_{ij} \approx 5.4$\,Å.
}
\label{f:L10_Jij_NN}
\end{figure}

\begin{figure*}[t]
\centering
\subfloat[]{%
\includegraphics[width=0.32\textwidth]{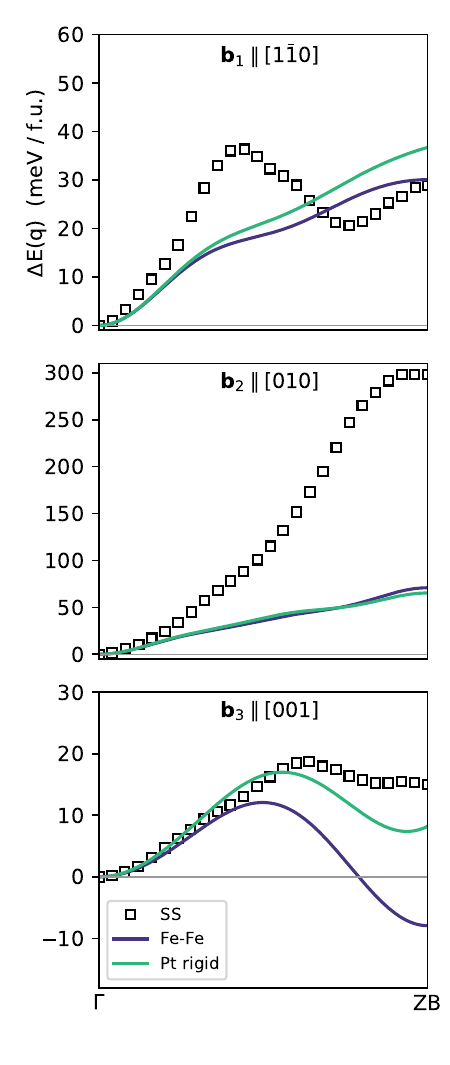}%
}\hfill
\subfloat[]{%
\includegraphics[width=0.32\textwidth]{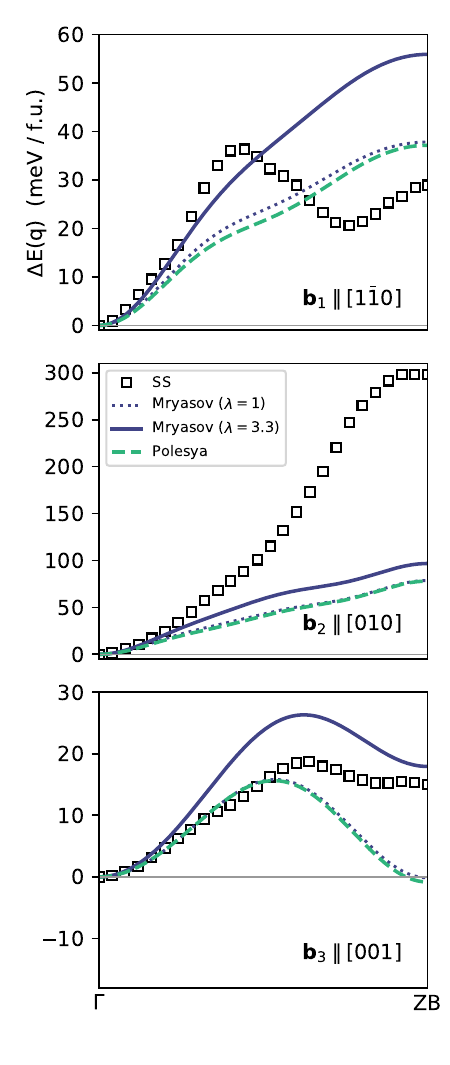}%
}\hfill
\subfloat[]{%
\includegraphics[width=0.32\textwidth]{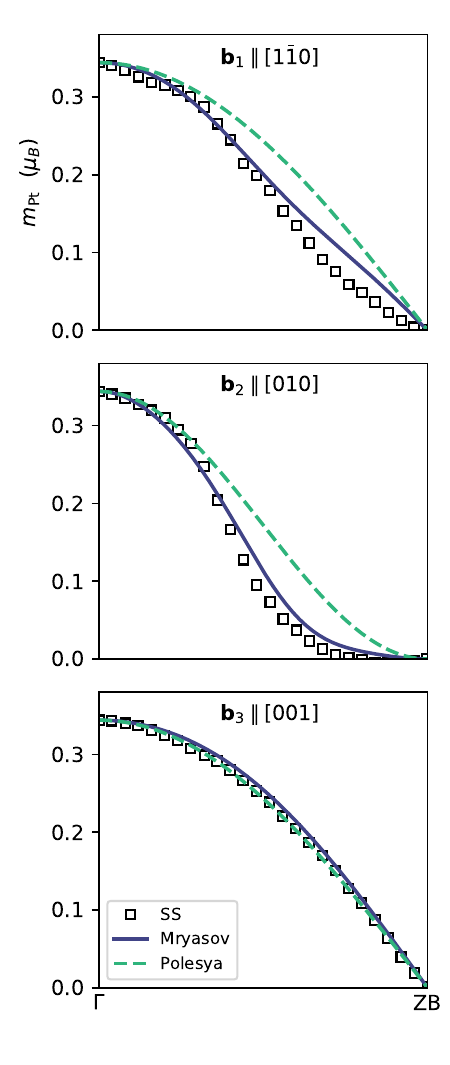}%
    }
\caption{Model comparison for FePt.
  (a) Spin-spiral energy $\Delta E(q)$ per formula unit along
  $\mathbf{b}_1$, $\mathbf{b}_2$, $\mathbf{b}_3$ (top to bottom),
  from the zone center $\Gamma$ to the zone boundary (ZB), for a
  planar ($\theta=90^{\circ}$) spiral. Open squares: self-consistent
  generalized-Bloch spin-spiral (SS) reference. Dark purple solid
  line: Fe--Fe-only model; green solid line: Pt rigid moment model,
  both built from the MFT couplings, Eq.\,(\ref{eq:dE_models}).
  (b) Same as (a), for the Mryasov model, Eq.\,(\ref{eq:mry_dE}),
  with amplitude $\lambda = 1$ (dark purple dotted line; the
  internally consistent value, Eq.\,(\ref{eq:mry_lambda})) and
  $\lambda = 3.3$ (dark purple solid line; midpoint of the range
  determined from the data in Fig.~\ref{fig:mry_fig1b}), and for the
  Polesya model, Eq.\,(\ref{eq:pol_dE}) (green dashed line).
  (c) Induced Pt moment magnitude $M_{\mathrm{Pt}}$ (in $\mu_B$)
  along the same directions as in (a) and (b). Open squares: SS;
  dark purple solid line: Mryasov model; green dashed line: Polesya
  model.
  }
\label{fig:FePt_models}
\end{figure*}

\subsection{Ground state from DFT}

As the first step, we take the experimental crystal structures of all L1$_0$ compounds and compute the total and atom-resolved magnetic moments; these data are summarized in the left half of Table~\ref{table:moments_TC}. The induced moments on Pt and Pd are very similar in magnitude for all studied systems, and the same applies to the orbital moments. Starting from a separate non-relativistic self-consistent calculation, we have computed the sum of eigenvalues for 3 orientations of the spin axis when the spin-orbit coupling is included in a one-shot calculation. The differences between these values allowed us to estimate the magnetic anisotropy constant $K_\mathrm{U}$, which is negative for all 4 systems, indicating an easy-axis anisotropy. Unsurprisingly, the FePt and CoPt systems have considerably larger $K_\mathrm{U}$ values than FePd and CoPd, most likely due to larger spin-orbit coupling effects induced by Pt compared to Pd. Also, by comparing the total energies of ferromagnetic and one selected antiferromagnetic configuration we have verified that the FM state is preferred for all compounds; the calculated energy differences $\Delta E_\mathrm{magn} = E_\mathrm{FM} - E_\mathrm{AFM}$ are listed as well in Table~\ref{table:moments_TC}. Interestingly, while the FM state clearly wins for FePd, CoPt and CoPd, as indicated by large energy differences, for FePt it is quite close to the AFM state where the spins in the neighboring Fe layers, spaced by Pt layers, are antiparallel. This coincides also with the result that will be discussed later on in this paper that the inclusion of Pt moments in the spin model is required for qualitatively correct description of magnetic ground state and Curie temperature.

\subsection{Computed $J_{ij}$ coupling parameters}

Fig.\,\ref{fig:Jij_all} shows the computed isotropic Heisenberg Fe-Fe coupling parameters $J_{ij}$ for all four studied systems. Both the spin spiral and MFT $J_{ij}$ are shown. The two approaches give qualitatively similar results, but it is clear that the differences between the two approaches are quite large. For instance, the nearest-neighbour $J_{ij}$ is more than twice as large in the spin-spiral approach compared to the MFT-approach for FePt. All four systems have a negative, i.e., AFM $J_{ij}$ coupling in the fifth Fe-Fe shell, which corresponds to the interlayer Fe-Fe exchange $J_{101}$ through Pt (Fig.~\ref{f:L10_Jij_NN}).

The AFM character of this coupling is in fact more pronounced in the spin-spiral approach compared to the MFT approach. We also see that its AFM character is largest for FePt and decreases as we traverse the series.

\subsection{Comparing models}
We will now compare a series of effective models with the spin-spiral reference. All of them are built from the same MFT exchange parameters evaluated at the ferromagnetic state and differ only in how they treat the induced Pt moment. We compare them by recomputing the spin-spiral energy $E(\mathbf{q})$ along the three high-symmetry directions $\mathbf{b}_1$, $\mathbf{b}_2$, $\mathbf{b}_3$, see Fig.~\ref{fig:FePt_models}(a,b), and through the induced Pt moment $m_{\mathrm{Pt}}(\mathbf{q})$, see Fig.~\ref{fig:FePt_models}(c). We first explain what each direction probes, then introduce the models.

The spin spirals with propagation vectors along $\mathbf{b}_1$, $\mathbf{b}_2$, and $\mathbf{b}_3$ probe complementary parts of the magnetic interaction network. Their energy depends on the phase factor $\mathbf{q}\cdot\mathbf{R}_{ij}$ associated with each exchange bond $\mathbf{R}_{ij}$. Bonds satisfying $\mathbf{q}\cdot\mathbf{R}_{ij}=0$ contribute only a constant energy shift and therefore do not influence the dispersion along that direction. Each propagation vector is consequently most sensitive to a different subset of exchange interactions.

The $\mathbf{b}_1 \parallel [1\bar{1}0]$ direction is parallel to a nearest-neighbor Fe--Fe bond. The corresponding spin spiral therefore directly modulates the dominant in-plane nearest-neighbor exchange. At the M point, neighboring Fe moments along this bond are antiferromagnetically aligned.

The $\mathbf{b}_2 \parallel [010]$ direction provides a complementary probe of the in-plane exchange interactions. Unlike $\mathbf{b}_1$, it is not parallel to a nearest-neighbor Fe--Fe bond, so the spin spiral modulates several in-plane exchange paths with different phase factors simultaneously. 
Consequently, the dispersion along $\Gamma$--X reflects the coherent superposition of phase-weighted contributions from several in-plane exchange shells. Depending on the propagation direction, these contributions may interfere either constructively or destructively, making this dispersion direction particularly sensitive to the combined in-plane exchange network.
At the X point, neighbouring Fe moments along the $[010]$ direction are antiferromagnetically aligned.

Finally, the $\mathbf{b}_3 \parallel [001]$ direction probes the interlayer magnetic coupling. Here the spiral propagates along the stacking direction of alternating Fe and Pt layers, so that only exchange paths with a finite $z$-component acquire a phase. Purely in-plane interactions therefore contribute only a constant background, making the resulting dispersion primarily sensitive to the interlayer Fe--Fe exchange and to the role of the induced Pt moments in mediating this coupling.

\subsubsection{Spin spiral energies from MFT exchange coupling parameters}

Having obtained the Heisenberg parameters $J_{ij}$ from MFT
at the ferromagnetic reference state, we now ask how well they
reproduce the flat spin-spiral energetics computed self-consistently with the
generalized Bloch theorem. This is a practically
important question, since the MFT couplings follow from a single ground-state
calculation, whereas each spin-spiral point requires a separate self-consistent
non-collinear calculation. The MFT approach is computationally very efficient, but how accurate are the results? To analyze this issue, we evaluate the energy of the classical
Heisenberg model with the MFT couplings for the same planar ($\theta=90^{\circ}$)
spirals and the same paths $\mathbf{b}_1$, $\mathbf{b}_2$, $\mathbf{b}_3$.
The per-formula-unit energy of the $\theta = 90^{\circ}$ spiral with respect to
the energy of the ferromagnetic state is
\begin{equation}
  \begin{split}
  \Delta E(\mathbf{q})
   ={}& \tfrac{1}{2}\sum_{j} J^{\mathrm{FeFe}}_{0j}
        \big[1-\cos(\mathbf{q}\cdot\mathbf{R}_{j})\big] \\
      +&\,\,\,\, \sum_{j} J^{\mathrm{FePt}}_{0j}
        \big[1-\cos(\mathbf{q}\cdot\mathbf{R}_{j})\big] \\
      +& \tfrac{1}{2}\sum_{j} J^{\mathrm{PtPt}}_{0j}
        \big[1-\cos(\mathbf{q}\cdot\mathbf{R}_{j})\big],
  \end{split}
  \label{eq:dE_models}
\end{equation}
where each sum runs over the neighbours $j$ of the representative Fe atom (Fe--Fe,
Fe--Pt) or Pt atom (Pt--Pt), and $\mathbf{R}_{j}$ is the bond vector. 

\paragraph{Fe--Fe only.}
Retaining only the first term treats FePt as if the Pt were nonmagnetic. The resulting spiral energies along $\mathbf{b}_1$, $\mathbf{b}_2$, and $\mathbf{b}_3$ are shown as dark purple lines in the leftmost column in Fig.~\ref{fig:FePt_models}. 
Along $\mathbf{b}_1$ the overall energy scale of the spin-spiral data is reproduced, with a tiny overshoot at the zone boundary. This is consistent with the fact that this direction is primarily governed by the nearest-neighbor Fe-Fe interaction. However, the non-monotonic shape is not captured. Along $\mathbf{b}_2$ the model reproduces only a small fraction of the (very large) spin-spiral energy. This suggests that the coherent superposition of phase-weighted contributions is only partially captured at this level of approximation. The most revealing failure is along $\mathbf{b}_3$. 
For $\mathbf{q}\parallel[001]$ the phase $\mathbf{q}\cdot\mathbf{R}_{j}$ vanishes for every in-plane bond ($R_{j}^{z}=0$), so the dominant, ferromagnetic in-plane Fe--Fe exchange does not contribute at all. Instead, the $[001]$ dispersion is set entirely by the much weaker interlayer Fe--Fe couplings.
The MFT Fe--Fe-only model places the antiferromagnetic stacking around 8\,meV/f.u.\ below the ferromagnetic state, whereas the spin-spiral data places it
around 15\,meV/f.u.\ above. Thus, the sign of the zone-boundary energy is wrong, and an Fe-Fe AFM stacking is predicted.
This clearly illustrates that the ferromagnetic stacking of the real material is not stabilized by Fe--Fe MFT exchange alone.

\paragraph{Rigid Pt moment model.}
The next level of approximation restores Pt as a magnetic site, retaining all three terms of Eq.~(\ref{eq:dE_models}) with the Pt treated as a rigid spin whose magnitude is frozen at its ferromagnetic value $M^{\mathrm{FM}}_{\mathrm{Pt}}$, so that the FM-reference couplings $J^{\mathrm{FePt}}$ and $J^{\mathrm{PtPt}}$ apply unchanged at all $\mathbf{q}$. The relaxed direction of the Pt magnetic moment is set to be parallel to the exchange field from the surrounding Fe atoms. The resulting spin spiral energies are plotted as green curves in the leftmost column in Fig.~\ref{fig:FePt_models}. 
Along the $\mathbf{b}_1$ direction, including the Pt interaction terms raises the zone boundary energy creating an overshoot of around 10\,meV compared to the spin-spiral data. The non-monotonic shape of the spin-spiral curve is still not reproduced. Along the $\mathbf{b}_2$ direction, the relative difference compared to the Fe-Fe-only model is minute, but note that the energy scale along this direction is an order of magnitude larger than along $\mathbf{b}_1$. Thus, including the Pt interactions does not capture the physics along this direction. The most interesting difference occurs along the $\mathbf{b}_3$ direction. Here, the zone boundary energy changes sign from negative to positive, curing the earlier qualitative error. Thus, at this level of approximation, the Fe-Fe interlayer coupling is correctly described as FM. The undershoot at the ZB is around 7\,meV. 

\begin{figure}
\vspace{10pt}
{\centering
\includegraphics[width=0.43\textwidth]{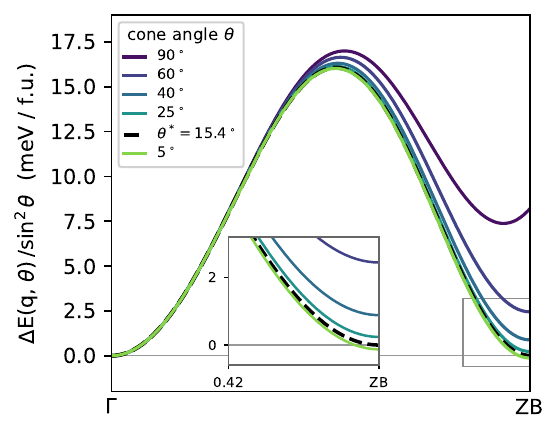}
}
\caption{Cone angle analysis of the Pt rigid moment model along the $\mathbf{b}_3$ direction. }
    \label{fig:ptrigid_coneangle}
\end{figure}

We now relax the $\theta = 90^{\circ}$ assumption to see if the FM configuration survives. It turns out that, indeed, this has profound effects at the $\mathbf{b}_3$ ZB.
Fig.\ref{fig:ptrigid_coneangle} shows the spin spiral energy along the $\mathbf{b}_3$ direction for a family of cone angles. For small enough cone angles, below approximately $15^{\circ}$, the AFM configuration becomes slightly energetically favorable.
(The equation for the spin spiral energy of the Pt rigid moment model for an arbitrary cone angle is given in Appendix~\,\ref{sec:cone_angle_formulas}.)

At a more general level, we note that the Pt-rigid model does not make much physical sense, since we hold the Pt moment magnitude at $M^{\mathrm{FM}}_{\mathrm{Pt}}$ for all
$\mathbf{q}$, thus over-counting the Pt moment contribution at the ZB where it has in fact collapsed. This illustrates a conceptual problem associated with MFT for systems with large longitudinal fluctuations in the magnetic moment magnitudes. 
Constructing spin Hamiltonians that allow us to relax the assumption of Pt rigid moments while still making use of the MFT-computed exchange couplings is therefore an interesting avenue. This leads us to the models developed by Mryasov \textit{et al.}\cite{Mryasov2005} and Polesya \textit{et al.}\cite{Polesya2010}, up next.

\subsubsection{Mryasov's model}
The two preceding models are the first members of a family of approximations that differ in their treatment of the induced Pt moment. The Fe--Fe-only model retains just the direct exchange between Fe moments, and the Pt rigid-moment model adds the Fe--Pt and Pt--Pt couplings while holding the Pt moment rigidly at its ferromagnetic value $M^{\mathrm{FM}}_{\mathrm{Pt}}$ for all $\mathbf{q}$, where $M^{\mathrm{FM}}_{\mathrm{Pt}}$ is the Pt moment magnitude in the ferromagnetic ground state ($M^{\mathrm{FM}}_{\mathrm{Pt}}\approx 0.34~\mu_{B}$). Both models misrepresent the Pt moment since in reality it changes its magnitude depending on the surrounding magnetic texture.
The model of Mryasov \textit{et al.}\cite{Mryasov2005}, originally introduced to describe the magnetocrystalline anisotropy of FePt and its temperature dependence, removes this deficiency by treating the Pt moment as a configuration-dependent degree of freedom. For the purpose of the present analysis, we retain only the isotropic (non-relativistic) terms of the Mryasov model, dropping the single-ion and two-ion anisotropy terms, which are spin--orbit derived and outside the scope of the present SOC-free analysis.

In the Mryasov model~\cite{Mryasov2005}, the Pt moment responds
linearly to the exchange field generated by the surrounding Fe atoms.
Mryasov \textit{et al.} introduce this linear response, together with
a quadratic Pt energy, as empirical inputs fitted to constrained-DFT
calculations; here we instead derive the model from an explicit spin
Hamiltonian, which makes its assumptions, its internal consistency
conditions, and the sign and magnitude of the resulting Pt-mediated
interaction transparent. The Hamiltonian has three distinct
contributions: a standard Heisenberg term describing the Fe--Fe
exchange, a term accounting for the Fe--Pt exchange interaction, and
a Pt intra-atomic contribution.
The Fe--Fe exchange term is described using
Eq.\,(\ref{eq:Hpairs}), i.e.,
\begin{equation}
   -\sum_{\langle i,j\rangle} J^{\mathrm{FeFe}}_{ij}\,
       \hat{\mathbf{e}}_i\cdot\hat{\mathbf{e}}_j .
\end{equation}
Likewise, the coupling between Fe and Pt is taken to be of the
general bilinear form (Eq.\,(\ref{eq:Hpairs}) in the Methods section). Explicitly,
this term becomes
\begin{equation}
  -\sum_{\nu}\mathbf{m}_{\mathrm{Pt},\nu} \cdot \mathbf{h}_{\nu},
  \qquad
  \mathbf{h}_{\nu} = \sum_{i} J^{\mathrm{FePt}}_{i\nu}\,
  \hat{\mathbf{e}}_{i},
  \label{eq:mry_coupling}
\end{equation}
where the sum over $\nu$ runs over the Pt sites,
$\mathbf{m}_{\mathrm{Pt},\nu}$ is the reduced Pt moment at site
$\nu$, and $\mathbf{h}_{\nu}$ is the exchange field at site $\nu$
created by the surrounding Fe atoms. The reduced Pt moment is
defined as $\mathbf{m}_{\mathrm{Pt},\nu} =
\mathbf{M}_{\mathrm{Pt},\nu}/M^{\mathrm{FM}}_{\mathrm{Pt}}$, where
$\mathbf{M}_{\mathrm{Pt},\nu}$ is the Pt magnetic moment at Pt site
$\nu$ and $M^{\mathrm{FM}}_{\mathrm{Pt}}$ is the Pt magnetic moment
magnitude at ferromagnetic order. Note that
$J^{\mathrm{FePt}}_{i\nu}$ is thereby defined as conjugate to the
reduced Pt moment, magnitude included, rather than to a unit vector
as in Eq.\,(\ref{eq:Hpairs}).
The third term, i.e., the on-site energy of the delocalized Pt
subsystem, is taken to have the form
\begin{equation}
C\sum_{\nu} \mathbf{m}_{\mathrm{Pt},\nu}^2 \, ,
\end{equation}
where $C$ is a positive constant. This form is dictated by general
considerations: in the absence of spin-orbit coupling the on-site
energy must be invariant under global spin rotations, hence a
function of $|\mathbf{m}_{\mathrm{Pt},\nu}|^{2}$ alone, and quadratic
order is the lowest allowed order.\cite{Marcus1988} Furthermore, Pt
does not order on its own, implying that $C$ must be positive, so
that the on-site energy has its minimum at
$\mathbf{m}_{\mathrm{Pt},\nu} = 0$.
Thus, the total spin Hamiltonian can be written as
\begin{equation}
    H =
    -\sum_{\langle i,j\rangle} J^{\mathrm{FeFe}}_{ij}\,
       \hat{\mathbf{e}}_i\cdot\hat{\mathbf{e}}_j
    -\sum_{\nu}\mathbf{m}_{\mathrm{Pt},\nu} \cdot \mathbf{h}_{\nu}
    + \,C\sum_{\nu} \mathbf{m}_{\mathrm{Pt},\nu}^2 \,.
    \label{Eq:Mryasov-Hamiltonian}
\end{equation}
The Pt moments $\mathbf{m}_{\mathrm{Pt},\nu}$ can be eliminated from
this expression as follows. Since the Pt moments are induced rather
than intrinsic, they carry no dynamics of their own but follow the
instantaneous Fe configuration adiabatically. For any fixed Fe
configuration $\{\hat{\mathbf{e}}_i\}$, the Hamiltonian is a strictly
convex quadratic function of each $\mathbf{m}_{\mathrm{Pt},\nu}$
separately, and the equilibrium Pt moments are therefore obtained by
minimization of $H$ with respect to $\mathbf{m}_{\mathrm{Pt},\nu}$,
which gives
\begin{equation}
  \mathbf{m}_{\mathrm{Pt},\nu} = \frac{1}{2C}\,\mathbf{h}_{\nu} .
  \label{eq:mry_slaving}
\end{equation}
The assumed linear response of the Pt moment is thus recovered as a
consequence of the quadratic Hamiltonian. Using
Eq.\,(\ref{eq:mry_slaving}), we can rewrite the last two terms in
Eq.\,(\ref{Eq:Mryasov-Hamiltonian}), per Pt site, as
\begin{equation}
  C\,\mathbf{m}_{\mathrm{Pt},\nu}^2
  - \mathbf{m}_{\mathrm{Pt},\nu}\cdot\mathbf{h}_{\nu}
  \,\Big|_{\mathbf{m}_{\mathrm{Pt},\nu} = \mathbf{h}_{\nu}/(2C)}
  = \frac{\mathbf{h}_{\nu}^{2}}{4C} - \frac{\mathbf{h}_{\nu}^{2}}{2C}
  = -\frac{\mathbf{h}_{\nu}^{2}}{4C} .
  \label{eq:mry_elimination}
\end{equation}
Thus, the intrinsic cost $\mathbf{h}_{\nu}^{2}/(4C)$ of polarizing
the Pt atom is overcompensated by the interatomic exchange energy
gain $\mathbf{h}_{\nu}^{2}/(2C)$, so the induced Pt moment lowers the
total energy. Next, we expand the sum over $\mathbf{h}_{\nu}^{2}$ to
obtain
\begin{equation}
  \begin{split}
  \sum_{\nu} \mathbf{h}_{\nu}^{2}
  &= \sum_{\nu}\sum_{i,j}
      J^{\mathrm{FePt}}_{i\nu} J^{\mathrm{FePt}}_{j\nu}\,
      \hat{\mathbf{e}}_i\cdot\hat{\mathbf{e}}_j \\
  &= 2\sum_{\langle i,j\rangle}
      \sum_{\nu}
      J^{\mathrm{FePt}}_{i\nu} J^{\mathrm{FePt}}_{j\nu}\,
      \hat{\mathbf{e}}_i\cdot\hat{\mathbf{e}}_j
    \;+\; E_0 ,
  \end{split}
\end{equation}
where $E_0 = \sum_{\nu}\sum_{i}
\big(J^{\mathrm{FePt}}_{i\nu}\big)^{2}$ arises from the $i=j$ terms.
$E_0$ is a constant, since
$\hat{\mathbf{e}}_i\cdot\hat{\mathbf{e}}_i = 1$, and can therefore be
omitted. The factor 2 appears since we go from a double sum to a sum
over unordered pairs.
We introduce the shorthand $J_{\mathrm{FePt}}\equiv\sum_{i}
J^{\mathrm{FePt}}_{i\nu}$ for the magnitude of the exchange field at
a Pt site in the ferromagnetic state; by the equivalence of all Pt
sites it is independent of $\nu$, so we suppress that index. The
Hamiltonian~(\ref{Eq:Mryasov-Hamiltonian}) then fixes its own
on-site constant: in the ferromagnetic state,
$\mathbf{h}_{\nu} = J_{\mathrm{FePt}}\,\hat{\boldsymbol{\eta}}$, and
Eq.\,(\ref{eq:mry_slaving}) gives
$|\mathbf{m}_{\mathrm{Pt},\nu}| = J_{\mathrm{FePt}}/(2C)$, which by
the normalization of the reduced moment must equal unity. Internal
consistency of the model therefore requires
$C = J_{\mathrm{FePt}}/2$. It is nevertheless useful to keep the
combination
\begin{equation}
  \lambda \equiv \frac{J_{\mathrm{FePt}}}{2C}
  \label{eq:mry_lambda}
\end{equation}
as an explicit dimensionless amplitude: within the model,
$\lambda = 1$, so when $\lambda$ is instead treated below as a free
amplitude and fitted to first-principles spin-spiral energies, its
deviation from unity quantifies physics not contained in
Eq.\,(\ref{Eq:Mryasov-Hamiltonian}). Collecting terms, we obtain, up
to the additive constant,
\begin{equation}
\begin{aligned}
H &=
-\sum_{\langle i,j\rangle}
\big(J^{\mathrm{FeFe}}_{ij}+\Delta J_{ij}\big)\,
\hat{\mathbf{e}}_{i}\cdot\hat{\mathbf{e}}_{j},\\
\Delta J_{ij}
&=
\frac{\lambda}{J_{\mathrm{FePt}}}
\sum_{\nu}
J^{\mathrm{FePt}}_{i\nu}J^{\mathrm{FePt}}_{j\nu} .
\end{aligned}
\label{eq:mry_Heff}
\end{equation}
The Pt-mediated correction $\Delta J_{ij}$ is built from products of
Fe--Pt couplings and therefore defines a positive-semidefinite
coupling matrix; in particular it is ferromagnetic whenever the
$J^{\mathrm{FePt}}_{i\nu}$ have a common sign, and its range is set
by twice the range of the Fe--Pt interaction, so that two Fe moments
sharing a polarizable Pt neighbor are coupled even in the absence of
a direct $J^{\mathrm{FeFe}}_{ij}$.

A value of $\lambda$ can also be determined directly from the
constrained-DFT results of Mryasov \textit{et al.} In Fig.~1(b) of
Ref.~\onlinecite{Mryasov2005}, the isotropic part of the energy
associated with the delocalized Pt states is given as a function of
the squared reduced Pt moment. We have digitized these data; they
are replotted in Fig.~\ref{fig:mry_fig1b} together with a
least-squares linear fit. The energy rises approximately linearly
from zero, identifying the plotted quantity as the polarization cost
of the induced Pt moment, measured from the unpolarized state. For a
linearly responding moment this cost equals
$\tfrac{1}{2}\,\mathbf{m}_{\mathrm{Pt},\nu}\cdot\mathbf{h}_{\nu}$
--- cf.\ the factor-of-two structure of
Eq.\,(\ref{eq:mry_elimination}) --- and its value in the
ferromagnetic state therefore coincides with the amplitude
$\lambda\,\tfrac{1}{2}J_{\mathrm{FePt}}$ of the Pt-mediated term in
the effective Hamiltonian. We stress that this identification relies
only on the linearity of the Pt response, not on the value of the
on-site constant $C$. The fitted slope is $1.83$~mRy per formula
unit, with an intercept of $0.10$~mRy/f.u.; allowing for the scatter
of the digitized points (the $m_{\mathrm{Pt}}^{2}=0$ point in
particular), slopes up to approximately $1.95$~mRy/f.u.\ are
compatible with the data. With
$\tfrac{1}{2}J_{\mathrm{FePt}} = 0.57$~mRy, this gives
$\lambda = 3.2\text{--}3.4$. Fitting the model instead directly to
our SS energies along $\mathbf{b}_3$ yields values of the same
order, though somewhat smaller: $\lambda = 2.3$ for a least-squares
fit over the full dispersion and $\lambda = 3.0$ when the
zone-boundary energy is matched (see
Fig.~\ref{fig:mryasov_lambdafit} in Appendix~\ref{sec:Mryasov_model}). All of these
determinations lie well above the value $\lambda = 1$ of the
internally consistent model.

\paragraph{Spiral energy.}
Since the Pt moments have been eliminated, the spiral energy follows
from Eq.\,(\ref{eq:mry_Heff}) alone and involves only the Fe degrees
of freedom. It is convenient to express the Pt-mediated part through
the normalized magnitude of the exchange field at a Pt site,
\begin{equation}
  \bar{h}(\mathbf{q})
  = \frac{1}{J_{\mathrm{FePt}}}
    \Big|\sum_{i} J^{\mathrm{FePt}}_{i\nu}\,
    \cos\!\big(\mathbf{q}\cdot\mathbf{R}_{i\nu}\big)\Big| ,
  \label{eq:mry_hbar}
\end{equation}
where $\mathbf{R}_{i\nu}$ connects the Pt site $\nu$ to the Fe site
$i$. The sine contributions cancel because the Fe environment of
each Pt site is centrosymmetric, $\bar{h}$ is independent of $\nu$
by the equivalence of the Pt sites, and $\bar{h}(0) = 1$ by
construction. The energy per formula unit relative to the
ferromagnetic state then becomes
\begin{equation}
  \begin{split}
  \Delta E(\mathbf{q})
   ={}& \tfrac{1}{2}\sum_{j} J^{\mathrm{FeFe}}_{0j}\,
         \big[1-\cos(\mathbf{q}\cdot\mathbf{R}_{j})\big] \\
      &+ \lambda\,\tfrac{1}{2}J_{\mathrm{FePt}}\,
         \big[1-\bar{h}^{2}(\mathbf{q})\big].
  \end{split}
  \label{eq:mry_dE}
\end{equation}
The first term is the direct Fe--Fe spiral energy, with the factor
$\tfrac{1}{2}$ assigning half of each bond to the representative Fe
site; the second is the induced-Pt contribution, whose factor
$\tfrac{1}{2}$ is inherited from Eq.\,(\ref{eq:mry_elimination}),
the net energy of a linearly induced moment being half its exchange
energy. The amplitude of the Pt term is set by $\lambda$, while its
$\mathbf{q}$-dependence is carried entirely by
$\bar{h}(\mathbf{q})$ and is therefore fixed by the Fe--Pt geometry.
Every ingredient except $\lambda$, i.e., $J^{\mathrm{FeFe}}_{0j}$,
$J^{\mathrm{FePt}}_{i\nu}$, and $J_{\mathrm{FePt}}$, is taken from
the computed MFT coupling parameters at the ferromagnetic reference
state. Discussion of numerical convergence of these quantities is in Appendix~\ref{si:mft_convergence}, Figs.~\ref{fig:jfept_convergence} and \ref{fig:fig_azb_b3_convergence}, while convergence of the spin-spiral energies is discussed in Appendix~\ref{sec:ss_convergence}, Figs.~\ref{fig:shell_cutoff_test}. Although the Pt moment no longer appears in
Eq.\,(\ref{eq:mry_dE}), $\bar{h}(\mathbf{q})$ retains a direct
physical counterpart: for a Pt moment responding linearly with the
observed unit slope it equals the reduced Pt moment of the spiral
state, and it indeed tracks the SS Pt moments closely,
see Fig.~\ref{fig:FePt_models}(c).

The admissible range of $\lambda$ is constrained by the structure of
the model itself. Consider any description in which (i) the Pt
moment is tied to the local exchange field,
$\mathbf{m}_{\mathrm{Pt},\nu} \parallel \mathbf{h}_{\nu}$ with
$|\mathbf{m}_{\mathrm{Pt},\nu}| \leq 1$, and (ii) the intra-atomic
energy is minimal at $\mathbf{m}_{\mathrm{Pt},\nu} = 0$ and
increases with $|\mathbf{m}_{\mathrm{Pt},\nu}|$ --- the defining
property of an induced moment. Relative to the ferromagnetic state,
the interatomic (Zeeman) contribution to the Pt energy of a spiral
cannot exceed $J_{\mathrm{FePt}}$, while the intra-atomic
contribution is necessarily \emph{negative}: the polarization cost
is largest in the ferromagnetic state, where the induced moment is
largest, and is reduced as the spiral quenches the
moment. The amplitude of the Pt term in Eq.\,(\ref{eq:mry_dE}) ---
its value where the exchange field is fully quenched,
$\bar{h} = 0$ --- therefore cannot exceed $J_{\mathrm{FePt}}$, i.e.,
\begin{equation}
  0 < \lambda \leq 2 ,
  \label{eq:mry_lambdabound}
\end{equation}
for the entire class of induced-moment models, with $\lambda = 1$ at
the internally consistent value $C = J_{\mathrm{FePt}}/2$ and
$\lambda \to 2$ only in the limit of a vanishing intra-atomic
energy. Note that this argument does not rely on the quadratic form
of the on-site term: the intra-atomic energy can only \emph{reduce}
the spiral amplitude below the bare exchange scale, never enhance
it.

The values of $\lambda$ determined above, ranging from $2.3$ to
$3.4$, all violate the bound~(\ref{eq:mry_lambdabound}).Clearly, the missing energy cannot be attributed to the Pt
intra-atomic term, nor to any refinement of it, within the
induced-moment framework. At the same time, the Pt \emph{moments}
themselves show no comparable enhancement --- the SS reduced Pt
moment follows $\bar{h}(\mathbf{q})$ closely,
Fig.~\ref{fig:FePt_models}(c) --- so the local response inferred
from the moments is unrenormalized. What the fit enhances is
therefore neither the moment response nor the on-site stiffness, but
the kernel of the Pt-mediated interaction itself,
$\chi_{\mathrm{Pt}}\,|J^{\mathrm{FePt}}|^{2}$: the spiral energies
determine only this product, and $\lambda$ is its dressing factor
relative to the bare estimate built from the static local Pt
susceptibility and the MFT couplings.

Requiring consistency with the
unrenormalized moments assigns the dressing to the coupling vertex:
an effective Fe--Pt interaction $\lambda\,J^{\mathrm{FePt}}_{i\nu}$
leaves the induced moments unchanged --- the susceptibility
parameter entering the induced-moment relation is rescaled accordingly,
since moments and energetics together determine only the product
$\chi_{\mathrm{Pt}}\,|J^{\mathrm{FePt}}|^{2}$ --- while enhancing
the mediated exchange by $\lambda$. We stress that this bookkeeping
does not diminish the physical Pt susceptibility: its strong
exchange enhancement is precisely the origin of the dressing, and
the decomposition only fixes how the enhancement is distributed
between the two factors of the kernel. In the language of the
Introduction, $\lambda$ thus measures the ratio of the dressed to
the bare Pt-mediated interaction.

At the level of the random-phase
approximation this ratio is expected to be of the order of the
Stoner enhancement factor of the induced-moment species,
$S = [1 - I\,N(\varepsilon_F)]^{-1}$, where $I$ is the Stoner
parameter and $N(\varepsilon_F)$ the one-spin Pt density of states
at the Fermi level; the fitted $\lambda \approx 3.3$ corresponds to
$I\,N(\varepsilon_F) \approx 0.7$, in line with the strong exchange
enhancement of Pt. The correspondence is approximate rather than
exact, most importantly because the enhancement inherits the
$\mathbf{q}$ dependence of the underlying susceptibility (see
below).

This is the Fe--Pt--Fe analogue of the distinction
between bare (magnetic-force-theorem) and renormalized exchange
parameters,\cite{Bruno2003,Katsnelson2004} and it parallels the
enhancement found in the Fe--Fe channel when the MFT couplings are
compared with those extracted from the full spin-spiral energies
(Fig.~\ref{fig:Jij_all}). A single constant $\lambda$ remains, of
course, a $\mathbf{q}$-averaged description: the structure in
$\Delta E(\mathbf{q})$ along $\mathbf{b}_1$ that no constant
amplitude reproduces (Figs.~\ref{fig:FePt_models} and
\ref{fig:mryasov_lambdafit}) is the direct signature of the
$\mathbf{q}$ dependence of $\chi_{\mathrm{Pt}}$, to which we return
in the Discussion (Sec.~\ref{sec:discussion}). 
This construction is practical, because all
couplings derive from a single MFT calculation at the ferromagnetic
reference, and the one remaining parameter, $\lambda$, can be fixed
by a handful of constrained-moment total-energy calculations, meaning that extensive spin-spiral calculations are not required. The Mryasov model thus turns the
computationally efficient fixed-reference MFT couplings into a
configuration-dependent energy, recovering much of the computationally expensive
spin-spiral (SS) physics at a small fraction of its cost. Here, where
the full SS reference is available, we have used it to test the
model rather than merely to parameterize it (see SI Sec.~III and Fig.~\ref{fig:mryasov_lambdafit}).

The resulting SS energies are plotted in the middle column of Fig.\,\ref{fig:FePt_models}. The dark purple line shows the result for $\lambda = 3.3$, the midpoint of the range determined above, whereas the dark purple dotted line uses $\lambda = 1$.
Along the $\mathbf{b}_1$ direction, Mryasov's model with $\lambda = 3.3$ overestimates the energy substantially, while along $\mathbf{b}_2$ it underestimates the energy significantly but less than the ``Fe-Fe only'' and ``Pt rigid moment'' models.
Along the $\mathbf{b}_3$ direction, it overestimates the zone-boundary (ZB) energy by approximately 3\,meV, but the model stabilizes the FM order along the $c$-direction with a margin of around 18\,meV.
As expected, Mryasov's model with $\lambda = 1$ looks qualitatively like an intermediate improvement step between the ``Fe-Fe only'' and ``Pt rigid moment'' models. The most interesting feature is again along the $\mathbf{b}_3$ direction, where this model with $\lambda = 1$ fails to clearly stabilize the FM order.

\subsubsection{Polesya's model}
An alternative configuration-dependent treatment of the induced
moment was proposed by Polesya \textit{et al.}\cite{Polesya2010} for
Fe$_x$Pd$_{1-x}$ and Co$_x$Pt$_{1-x}$ alloys, and can be transferred
directly to the FePt case. In contrast to the Mryasov's model, in
which the Pt moments are eliminated analytically, the Polesya's model
retains the Pt as an explicit magnetic site in an extended
Heisenberg Hamiltonian [Eq.~(1) in Ref.~\onlinecite{Polesya2010}],
\begin{equation}
  \begin{split}
  H = {}& -\sum_{\langle i,j\rangle} J^{\mathrm{FeFe}}_{ij}\,
            \hat{\mathbf{e}}_i\cdot\hat{\mathbf{e}}_j
         -\sum_{i,\nu} J^{\mathrm{FePt}}_{i\nu}\,
            \hat{\mathbf{e}}_i\cdot\mathbf{s}_\nu \\
        & -\sum_{\langle \nu,\nu'\rangle} J^{\mathrm{PtPt}}_{\nu\nu'}\,
            \mathbf{s}_\nu\cdot\mathbf{s}_{\nu'} \, ,
  \end{split}
  \label{eq:pol_H}
\end{equation}
where $\mathbf{s}_\nu$ is the reduced Pt moment at site $\nu$, in
the same convention as in Eq.\,(\ref{eq:mry_coupling}): the
couplings are conjugate to the reduced moment and include the actual magnitude of Pt moments
[in Ref.~\onlinecite{Polesya2010} this normalization is made
explicit in Eq.~(5)]. The Pt terms therefore scale with the
induced amplitude, linearly for Fe--Pt and quadratically for
Pt--Pt.

Instead of being obtained by energy minimization, the Pt moment is
prescribed: Polesya \textit{et al.} take it proportional to the
vector sum of the nearest-neighbour Fe moments, with the
proportionality constant fixed by the ferromagnetic ground state
[Eqs.~(2)--(3) in Ref.~\onlinecite{Polesya2010}]. Since the Fe moment magnitude is
configuration-independent, this prescription reads, in reduced
units,
\begin{equation}
  \mathbf{s}_\nu = \frac{1}{N_{\mathrm{NN}}}
    \sum_{j\in\mathrm{NN}(\nu)} \hat{\mathbf{e}}_j \, ,
  \label{eq:pol_moment}
\end{equation}
where the sum runs over the $N_{\mathrm{NN}} = 8$ nearest Fe
neighbours of the Pt site. For a flat spiral these neighbours carry
phases $\mathbf{q}\cdot\mathbf{R}_{\nu j}$, and
Eq.\,(\ref{eq:pol_moment}) gives
$\mathbf{s}_\nu = s(\mathbf{q})\,\hat{\mathbf{u}}_\nu$, where
$\hat{\mathbf{u}}_\nu$ is the spiral direction at the Pt position
and
\begin{equation}
  s(\mathbf{q})
  = \frac{1}{N_{\mathrm{NN}}}\sum_{j\in\mathrm{NN}}
      \cos(\mathbf{q}\cdot\mathbf{R}_{\nu j}) \, ,
  \label{eq:pol_sq}
\end{equation}
a signed amplitude ($s$ may become negative when the neighbour sum
reverses). This differs from the Mryasov model in several ways: the
response is restricted to the first shell and is unweighted by the
exchange couplings, and the model contains, by construction, no
intra-atomic term and no adjustable amplitude --- it operates at the
undressed level, $\lambda = 1$, of the previous subsection.

Inserting the spiral configuration into Eq.\,(\ref{eq:pol_H}), the
energy per formula unit relative to the ferromagnetic state is
\begin{equation}
  \begin{split}
  \Delta E(\mathbf{q})
   ={}& \tfrac{1}{2}\sum_{j} J^{\mathrm{FeFe}}_{0j}
          \big[1-\cos(\mathbf{q}\cdot\mathbf{R}_j)\big] \\
      &+ \sum_{j} J^{\mathrm{FePt}}_{0j}
          \big[1-s(\mathbf{q})\cos(\mathbf{q}\cdot\mathbf{R}_{0j})\big] \\
      &+ \tfrac{1}{2}\sum_{j} J^{\mathrm{PtPt}}_{0j}
          \big[1-s^{2}(\mathbf{q})\cos(\mathbf{q}\cdot\mathbf{R}_{j})\big] ,
  \end{split}
  \label{eq:pol_dE}
\end{equation}
where $j$ runs over the Fe neighbours of the representative Fe site
in the first term, its Pt neighbours in the second, and the Pt
neighbours of the representative Pt site in the third. This is
reminiscent of the Pt-rigid-moment model discussed earlier, but with
the Pt amplitude replaced by the configuration-dependent
$s(\mathbf{q})$; at $s = 1$, Eq.\,(\ref{eq:pol_dE}) reduces to the
rigid model.

The green dashed lines in the middle column of Fig.\,\ref{fig:FePt_models} show the Polesya model spin spiral energies. We conclude that the model gives results very similar to the Mryasov model for $\lambda = 1$.
This close agreement is no accident: the intra-atomic Pt energy of
the Mryasov model is present in the Polesya model in disguise. The
Pt--Pt couplings sum to a large antiferromagnetic value,
$\sum_j J^{\mathrm{PtPt}}_{0j} = -16.9$~meV, so that for slowly
varying textures ($\mathbf{s}_\nu \approx \mathbf{s}_{\nu'}$) the
last term of Eq.\,(\ref{eq:pol_H}) contributes
$+\tfrac{1}{2}\big|\sum_j J^{\mathrm{PtPt}}_{0j}\big|\,s^{2}$ per Pt
site. Since, in addition, $s(\mathbf{q}) \approx
\bar{h}(\mathbf{q})$ (the nearest-neighbour shell carries most of
$J_{\mathrm{FePt}}$), the two models evaluate nearly the same energy
functional at nearly the same induced moment. 

The rightmost column in Fig.\,\ref{fig:FePt_models} shows the Pt magnetic moment magnitude along the 
$\mathbf{b}_1$, $\mathbf{b}_2$ and $\mathbf{b}_3$ directions. Both the Mryasov and Polesya models capture the collapse of the Pt moment as $\mathbf{q}$ goes from the zone center to the zone edge, with the Mryasov model reproducing the shapes of the curves more precisely.  

\subsubsection{Pt rigid moment and the Mryasov models for small cone angle}
\begin{figure}
\vspace{10pt}
{\centering
\includegraphics[width=0.43\textwidth]{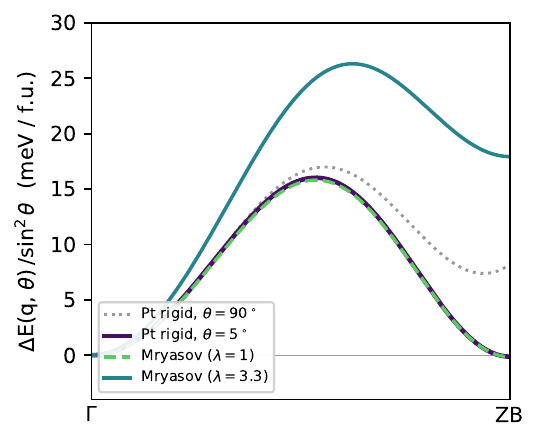}
}
\caption{Small cone angle comparison of the ``Pt rigid moment'' and Mryasov's models along the $\mathbf{b}_3$ direction.}
\label{fig:ptrigid_vs_mryasov_smallcone}
\end{figure}

In Fig.\,\ref{fig:ptrigid_vs_mryasov_smallcone} we compare the ``Pt rigid moment'' model with the Mryasov's model for a small cone angle, $\theta = 5^{\circ}$. In this limit, we find that the ``Pt rigid moment'' model corresponds to the Mryasov's model with $\lambda = 1$. 

\subsection{Critical temperatures, magnon spectra, spin stiffness}

The models above were all evaluated against the computed spin-spiral energy, which characterizes the static, zero-temperature energy landscape. 
In the end, what we are really interested in is being able to describe observable properties. We therefore now present results for critical temperatures and the dynamical structure factors computed using Monte Carlo and ASD simulations and two data sets: the MFT Fe-Fe only exchange parameters and the full spin-spiral spin Hamiltonian with cone angle $\theta = 90^{\circ}$.

\subsubsection{Critical temperatures}

\begin{figure}
{\centering
\includegraphics[width=0.48\textwidth]{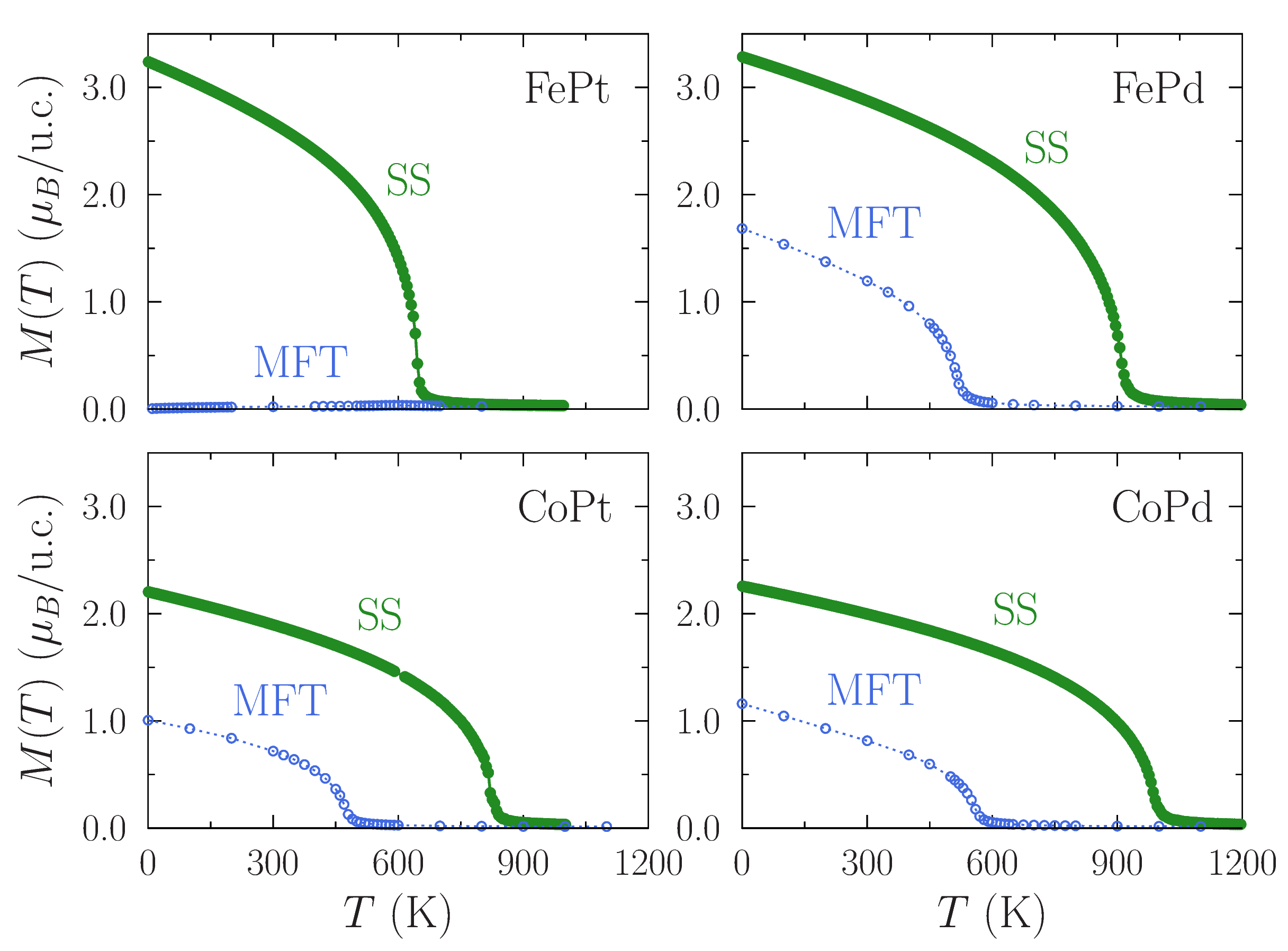}
}
\vspace{-20pt}
\caption{
Temperature-dependence of the magnetization calculated using Monte Carlo simulations based on \textit{ab initio} Heisenberg interaction parameters and uniaxial anisotropy. The Heisenberg parameters are obtained using the magnetic force theorem (MFT) (blue dashed lines, open symbols) or Fourier transform of the generalized Bloch spin-spiral energies (green lines, filled symbols), same as in Fig.~\ref{fig:Jij_all}.
}
\label{fig:MT_all}
\end{figure}

Computed critical temperatures are listed in Table\,\ref{table:moments_TC}.
$T_{\mathrm{C}}^{\mathrm{MFT}}$ (third column from the right) denotes the critical temperature computed using the MFT Fe-Fe only exchange parameters. Note that from the temperature-dependent magnetizations in Fig.\ref{fig:MT_all}d-f it is only possible to get a rough estimate of the ordering temperatures. We obtained the precise values reported here from the fourth-order cumulant method \cite{binder2012monte}.

For FePt, the simulation yields an AFM ground state, in agreement with our analysis above: for the Fe-Fe only model, the direct interlayer Fe--Fe coupling along the $\mathbf{b}_3$ direction is AFM, and the Pt-mediated Fe--Pt--Fe channel that restores the FM stacking is absent.
In contrast, for FePd, CoPt, and CoPd, the MFT-based model yields realistic critical temperatures, albeit on the low side compared to experimental data (listed in the right-most column).
The spin-spiral exchange parameters systematically yield critical temperatures ($T_{\mathrm{C}}^{\mathrm{SS}}$, second column from the right) in closer agreement with experiment. For FePt, the correct physics is recovered since the FM state is predicted as the ground state, showcasing the qualitative difference between the MFT and the SS spin Hamiltonian descriptions for this system.

The simulated ordering temperatures from the SS-derived exchange parameters agree rather well with experimental observations. For example, $750\, \mbox{K}$ was reported as an experimental Curie temperature for FePt\cite{LYUBINA2011291,200601904,klemmer1995magnetic}, while our theoretical prediction is 640\,K (see Fig.~\ref{fig:MT_all}d). If exchange parameters are derived from spin-spiral calculations at $\theta =45^{\circ}$ instead of $90^{\circ}$, the Curie temperature for FePt is around 680\,K, making it even closer to experimental observations. There is a physical intuition for this, since at the Curie temperature the magnetic energy is roughly half its zero-temperature value, suggesting that the average angle between the fluctuating Fe moments and total magnetization is around $\theta =45^{\circ}$. More detailed discussion of how the spin-spiral cone angle changes the calculated spin model can be found in Appendix~\ref{sec:cone_angle-magnons}, where this issue is discussed in the context of spin-spiral energies (Fig.~\ref{fig:spirals_different_theta}), effective Fe-Fe interactions (Fig.~\ref{fig:Jij_different_theta}), and adiabatic magnon spectra (Fig.~\ref{fig:ams_different_theta}).

Concerning the other systems, For FePd a measured Curie temperature of $T_\mathrm{C}=734-760\,\mbox{K}$ was reported\cite{weller2000high,kozubski2022atomic,wang2004effects,klemmer1995magnetic} which is comparable to our theore\-tical result (906\,K). Experimental data for bulk CoPd are scarce \cite{matsuo1972ordered}, with some works reporting that CoPd can be metastable in the L1$_0$ structure\cite{YABUHARA20118359}. 
Most reports on the magnetic properties refer to thin films or disordered systems of CoPd and report ordering temperatures in a wide range between $650-1293\,\mbox{K}$ for Co concentrations from $30\%$ to $84\%$ \cite{PhysRev.125.490,ungemach1980temperature}. 

We should note that direct comparison of critical temperatures calculated from classical spin models with experiment requires some attention, which was noticed already in one of the first papers on the magnetic force theorem [\onlinecite{LiechtensteinCurie1985}]. Quantum effects should increase the theoretical value of $T_c$; in the mean-field approximation for quantum Heisenberg model, the corresponding factor is $1 + 1/S$ which is not included in our consideration. Nevertheless, based on comparison of the whole set of experimental and theoretical data we can say that the spin-spiral results are the most reliable.

\subsubsection{Magnon spectra}

\begin{figure}[htb]
\includegraphics[width=0.42\textwidth]{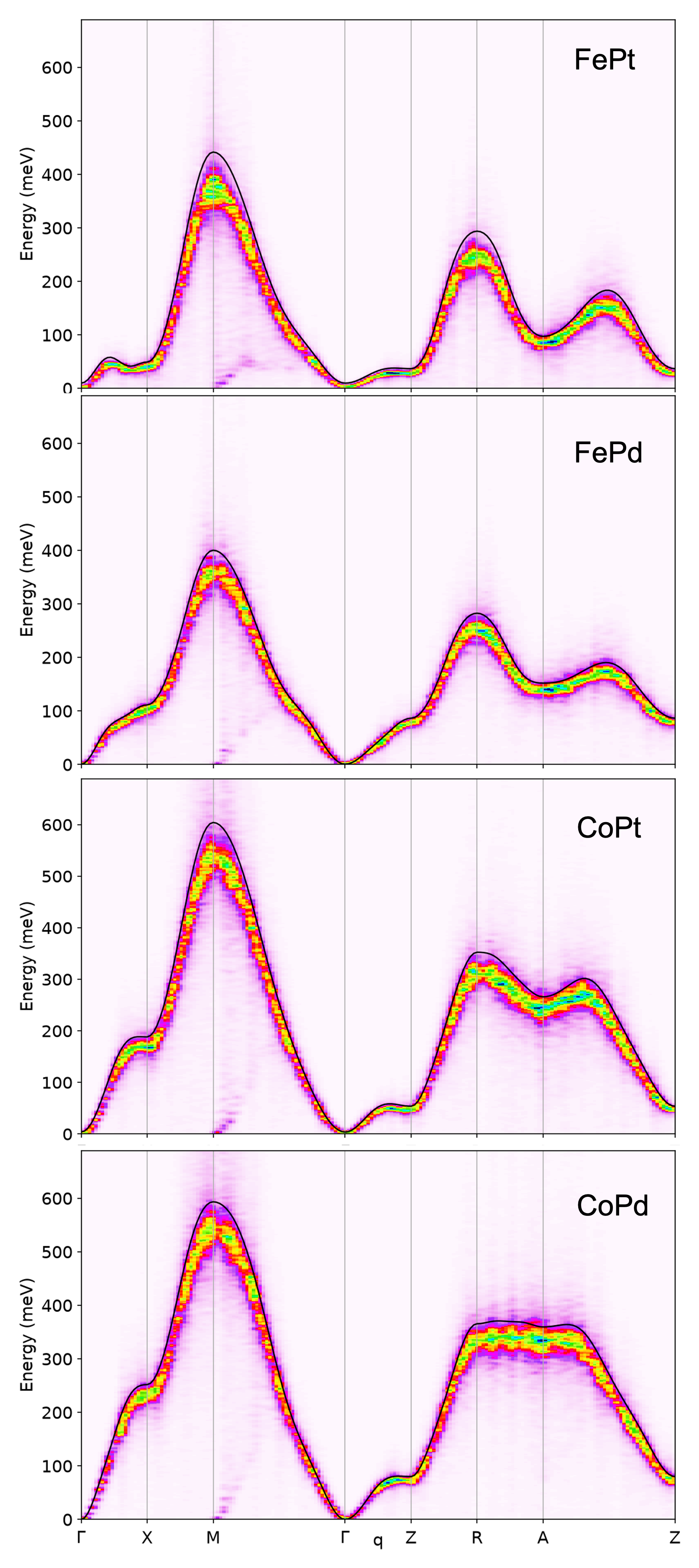}
\caption{Spin-spin dynamic structure factor $S^k(\mathbf{q},w)$ (color shading) at $T=300 \,\mbox{K} $ with a Gilbert damping $\alpha=0.044$.
Adiabatic magnon spectra (black solid lines) calculated at $T=0\,\mbox{K}$.}
\label{fig:magnon_spectra}
\end{figure}

In Fig.~\ref{fig:magnon_spectra} we present adiabatic magnon spectra based on the $J_{ij}$ parameters computed in the spin-spiral approach, for FePt, FePd, CoPt, and CoPd, calculated using atomistic spin dynamics at $T=0\,\mbox{K}$ and linear spin wave theory \cite{eriksson2017atomistic,etz2015atomistic}. These two methods provide the spin-spin dynamic structure factor $S^k(\textbf{q},\omega)$ and adiabatic magnon spectra, respectively, which are compared in the figure.

At zero temperature both methods give the same results. The impact of temperature  on the structure factor is illustrated in Fig.~\ref{fig:magnon_spectra}, where simulations were performed for a temperature of $T=300 \,\mbox{K}$. Here, we have used a realistic value of Gilbert damping $\alpha=0.044$ (which is within a range reported in Ref. \onlinecite{sasaki2023thermal} for FePt, and in the same order as reported in Ref.~\onlinecite{kawai2014gilbert} for FePd). The same damping constant is used for all studied materials due to the limited amount of experimental data. For FePd, the Gilbert damping is reported to be around 0.01 \cite{peria2024low,PhysRevMaterials.6.113402}. Increasing the temperature from 0\,K to 300\,K leads to a slight softening of magnons for all four studied systems, as indicated by the black solid lines in Fig.\ref{fig:magnon_spectra} which are just above the color-shaded regions corresponding to the finite-temperature structure factor. Note that the chosen temperature ($T=300\,\mbox{K}$) is quite low in comparison to the respective ordering temperatures of each system.

Unfortunately, there is currently limited experimental data, which impedes direct benchmarking of our calculations, leaving the results presented here as a prediction. However, in Ref.~\onlinecite{qin2015long}, a chemically-disordered 2D face-centered tetragonal alloy film of FePd was studied, and the magnon dispersion was measured. To compare with this experiment, simulations for two atomic layers were performed with reasonable agreement between theory and experiment (data not shown). These simulations and the experimental data show a softening of the magnon energies compared to reported bulk magnons, as was also reported in Ref. \onlinecite{etz2015atomistic}, and the agreement between theory and experiment illustrates that exchange parameters of FePd-based systems can reproduce experimental magnons. Since the level of complexity of the electronic structure and magnetism is expected to be very similar for the systems studied here, we anticipate that the theoretically predicted magnons shown in Fig. \ref{fig:magnon_spectra} should be reasonably accurate in comparison to experiment.

\subsubsection{Spin stiffness}

\begin{figure}[h]
{\centering
\includegraphics[width=0.48\textwidth]{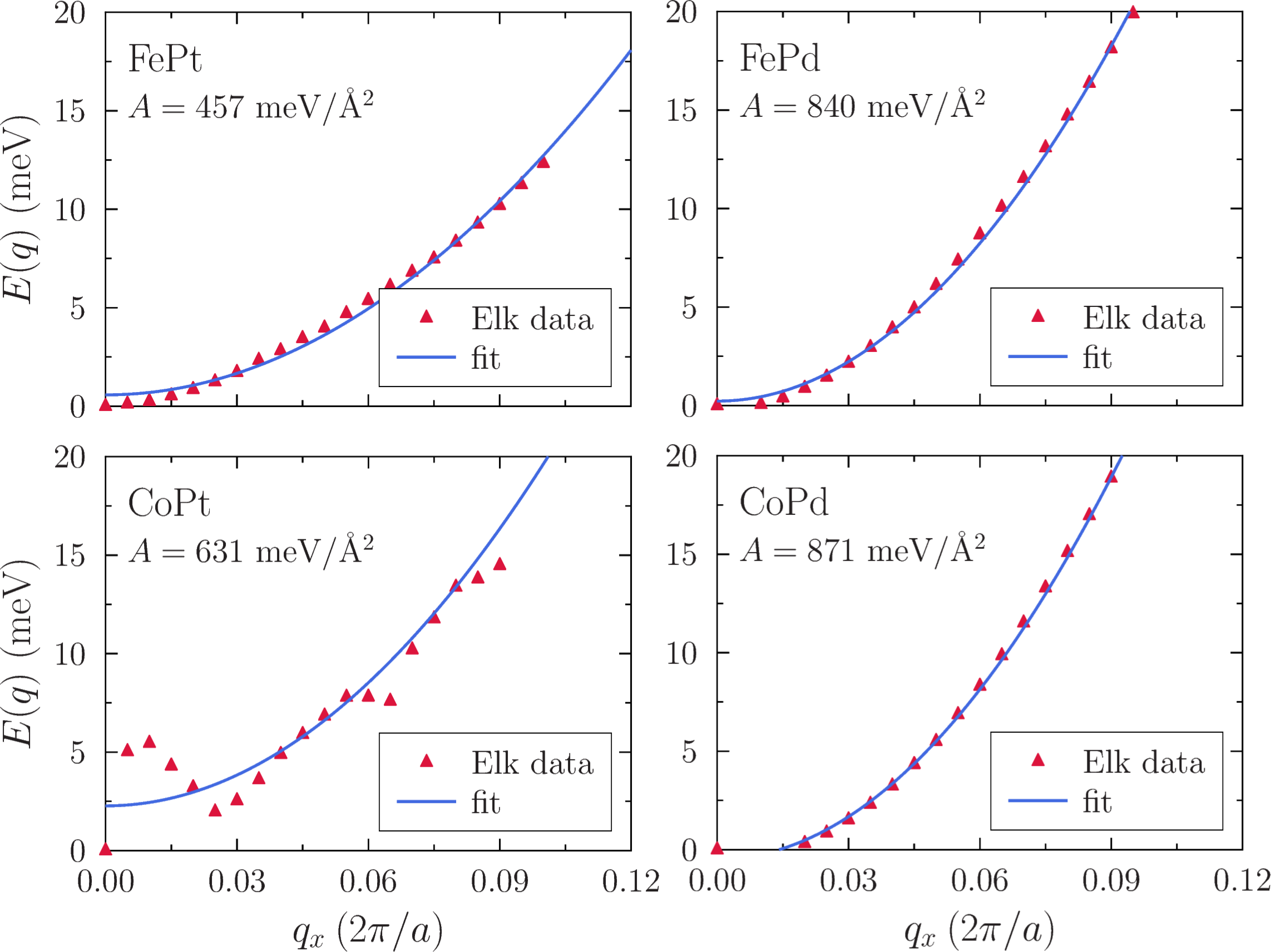}
}
\vspace{-15pt}
\caption{Spin-spiral energies $E(\vec{q})$ calculated along the [100] direction around the $\Gamma$ point plotted together with the quadratic fit functions to determine the spin stiffness constant $A$.}
\label{f:Eq_from_Elk}
\end{figure}

For reference, we provide here the estimates of the spin stiffness obtained from the spin-spiral energy curvature near the Gamma point and from the summation of exchange interactions in real space (see details e.g.~in [\onlinecite{Borisov2024rev}]). These results are shown in Table~\ref{table:stiffness} and Fig.~\ref{f:Eq_from_Elk}. Overall, the order of magnitude of the spin stiffness is the same for all systems. The fact that the spin stiffness estimates from the spin-spiral and rigid-moment data are not identical is not surprising: this is, in fact, related to the discussion about the dressed susceptibility in the next section and the conclusions.

\setlength{\tabcolsep}{5pt}
\begin{table}
  \caption{Spin stiffness (in units of $\mathrm{meV}\cdot\mathrm{\AA}^2$) calculated from the spin-spiral energy curvature (Fig.~\ref{f:Eq_from_Elk}) and real-space summation of Heisenberg exchange interactions in the spin model with rigid Fe and Pt moments (Fig.~\ref{f:all_Jijs}); three diagonal components of the spin stiffness tensor are given as well.}
  \centering
  \vspace{5pt}
  \begin{tabular}{c|c|cccc}
    \hline
    Compound & Spin spirals & \multicolumn{4}{c}{Rigid-moment spin model} \\
     &  & $A_\mathrm{total}$ & $A_{xx}$ & $A_{yy}$ & $A_{zz}$ \\
    \hline
      FePt &  457  & 462 & 283 &  94 &  88 \\
      FePd &  840  & 675 & 425 & 127 & 125 \\
      CoPt &  631  & 504 & 267 &  94 & 154 \\
      CoPd &  871  & 619 & 309 & 103 & 128 \\
    \hline
  \end{tabular}
  \label{table:stiffness}
\end{table}

\section{Discussion}
\label{sec:discussion}

The comparisons presented above allow for the large discrepancy between the MFT and SS exchange parameters --- and the qualitative failure of the MFT description for FePt --- to be discussed from a unified perspective. We first note that the discrepancy is due to a physical reason, not a particular numerical implementation of the MFT approach: as shown in Appendix~\ref{sec:emto} and Fig.~\ref{fig:compare_emto_rspt}, two independent
electronic-structure methods produce very similar MFT interaction parameters.

The family of models compared in the Results section identifies the physical reason for the aforementioned discrepancy between the MFT and SS parameters.
Each model can be read as a different approximation of the response
of the induced Pt moments. Without the Pt interaction channel (i.e.~in a model with Fe--Fe interactions only),
the predicted interlayer ground state is qualitatively wrong. Freezing
the Pt moment at its ferromagnetic magnitude (``Pt rigid moment'' model)
restores the ferromagnetic stacking for planar spirals, but the
cone-angle analysis exposes this as fragile --- the ferromagnetic
state survives only above $\theta^{*}\!\approx\!15^{\circ}$, and in
the small-cone-angle limit the rigid model coincides with the Mryasov's model
at $\lambda = 1$. That level of theory --- the static local Pt susceptibility
combined with the bare MFT couplings, shared by the Mryasov's
($\lambda = 1$) and Polesya's models --- leaves the interlayer
ferromagnetism marginally stable. Reproducing the \textit{ab initio} spin-spiral energies requires the
amplitude $\lambda \approx 3.3$, and the bound of
Eq.~(\ref{eq:mry_lambdabound}) turns this number into a statement of
mechanism: no induced-moment model with an intra-atomic Pt energy can provide that value of $\lambda$, and the non-renormalized Pt moments
(Fig.~\ref{fig:FePt_models}(c)) rule out the moment response.
The enhancement must therefore reside in the Pt-mediated interaction
kernel $\chi_{\mathrm{Pt}}|J^{\mathrm{FePt}}|^{2}$ --- in the language
of the Introduction, the magnetic energies are governed by the dressed
susceptibility, of which the bare response probed by the
frozen-potential MFT captures only about a third. 

This is the central
result of the present work: for induced-moment systems the
bare-versus-dressed distinction is not a quantitative correction but, in the case of FePt, 
decides the magnetic ground state.

The residual discrepancies delineate what no single-amplitude
bilinear spin model can capture. Along $\mathbf{b}_1$ the SS dispersion is
non-monotonic --- a \emph{shape}, which no constant $\lambda$ can
reproduce and which points to the $\mathbf{q}$ dependence of
$\chi_{\mathrm{Pt}}$. Along $\mathbf{b}_2$ a large shortfall in
\emph{magnitude} remains; since the Fe moment magnitudes are
constrained to their ferromagnetic values in the SS reference, this
missing energy is transverse, multi-shell interference physics rather
than longitudinal Fe fluctuations, and remains an open question.

Also, probably the most direct way to derive the MFT exchange parameters is suggested in Ref.[\onlinecite{Katsnelson2000}]. As a first step, one uses the magnetic force theorem, which is formally exact in both density functionals [\onlinecite{LKAG1987}] and in Green's function functionals [\onlinecite{Katsnelson2000}] and gives an explicit expression of the spin torque acting on a rotating spin on a given site in terms of  $\hat{G}$ and $\hat{\Delta}$. As the second step, this term is decomposed into the sum of pairwise terms which are compared with those in the effective classical Heisenberg model. Although this procedure may look exact, it has a problem. Namely, in both density functional and Green's function functional the configuration with a deviated spin on one site cannot be considered self-consistently without adding a constrained magnetic field which should be added to the derivation and modify the expression for exchange parameters [\onlinecite{Bruno2003},\onlinecite{Solovyev2024},\onlinecite{Solovyev2021}].

The relation between MFT [\onlinecite{LKAG1987}] and the approach by Bruno [\onlinecite{Bruno2003}], in terms of inhomogeneous spin susceptibility, was discussed in Ref.[\onlinecite{Katsnelson2004}] (see also Ref.[\onlinecite{Szilva2023}]). It turns out that the former parameters describe the poles of dynamical susceptibility and, thus, magnon energies probed by inelastic neutron scattering whereas the latter ones correspond to the values of the static susceptibility and, thus, to the energies of the spin spiral. The difference is supposed to be small only in the regime of well-defined local magnetic moments when typical magnon energies are much smaller than the Stoner splitting which is, more or less, the case of \textit{bcc} Fe but not of \textit{fcc} Ni [\onlinecite{Katsnelson2004}]; more examples can be found in [\onlinecite{Solovyev2024},\onlinecite{Solovyev2021}]. For an analysis of the applicability of the MFT approach in terms of diagram techniques of quantum many-body theory see Refs.~[\onlinecite{Stepanov2018},\onlinecite{Stepanov2022}]. Physically, conditions of its applicability can be related again to the existence of well-defined local magnetic moments. Keeping in mind that Pt/Pd magnetic moments which play an important role for the system under consideration are obviously not well-localized it is not surprising that MFT and spin-spiral exchange parameters are so different in this case. Note that in such situations, spin-spiral exchange parameters are much more reliable when calculating the Curie temperature and other thermodynamic properties [\onlinecite{Katsnelson2004}], where large deviations from the magnetic ground state are relevant and must be included in the calculations. Our results for the Curie temperatures for the full L1$_0$ family (Table~\ref{table:moments_TC}) support this statement across four systems, and the amplitude $\lambda$ introduced above quantifies the same dichotomy --- pole-type (bare) versus static (renormalized) parameters --- specifically in the Fe--Pt--Fe channel.

Within the local spin-density approximation, the spin-wave stiffness calculated in both ways should coincide, but this can be proven rigorously only in the case of one magnetic atom per elementary cell [\onlinecite{Katsnelson2004}]; for our systems even this is not valid, since the unit cell contains at least one Fe and one Pt atom. The comparison of the stiffness constants extracted from the MFT and SS parameter sets (see Table~\ref{table:stiffness}) therefore provides an independent test of how far the two descriptions already deviate at leading order in $\mathbf{q}$, where the Pt-mediated ($\lambda$-enhanced) contribution enters the curvature of the magnon dispersion, in particular along $[001]$.

Our analysis deliberately excluded spin--orbit coupling. In the model of Mryasov \textit{et al.},\cite{Mryasov2005} the magnetocrystalline anisotropy of FePt is itself configuration dependent: the Pt easy-axis energy enters through the induced moment, and eliminating the slaved Pt degrees of freedom produces (alongside the isotropic mediated exchange) a two-ion anisotropy built on the same Fe--Pt--Fe product kernel. This configuration-dependent magnetic anisotropy acts alongside the isotropic Pt-mediated interaction and may work on the margin to stabilize the ferromagnetic order along $[001]$. \cite{Mryasov2004}
The scale of this effect is an order of magnitude smaller than the dressing deficit identified above, i.e.,  configuration-dependent anisotropy cannot repair the MFT ground-state failure in FePt.

Furthermore, disorder may exist in these systems. The direct Fe--Fe couplings in FePt were, in earlier work, found to be sensitive to chemical order,  potentially destabilizing the ferromagnetic state.\cite{Brown2003FePt}

Finally, we note that the slow convergence of the real-space exchange interactions (Appendices~\ref{si:mft_convergence} and~\ref{sec:ss_convergence}) should not primarily be interpreted as evidence for unusually long-ranged pairwise physics. Rather, it reflects the pronounced momentum dependence of the underlying electronic susceptibility. The exchange parameters are the Fourier coefficients of this susceptibility-mediated interaction, so a strongly structured $J(\mathbf{q})$ necessarily requires many coordination shells for an accurate real-space representation. In this sense, the large number of pair interactions is a consequence of approximating a fundamentally nonlocal electronic response within a local real-space Hamiltonian --- the same nonlocality that appears as the $\mathbf{q}$ dependence of $\chi_{\mathrm{Pt}}$ above.

\section{Conclusions}

In this theoretical work, we have compared different ways of
constructing effective spin Hamiltonians for the ordered L1$_0$
compounds FePt, FePd, CoPt, and CoPd, in which the key difference is
the treatment of the induced moments on Pt and Pd. Conventional
magnetic-force-theorem (MFT) parameters, evaluated at the
ferromagnetic reference state, fail qualitatively for FePt ---
predicting an antiferromagnetic interlayer ground state --- and
underestimate the ordering temperatures of the other three
compounds, whereas parameters extracted from self-consistent
spin-spiral calculations describe the magnetic properties well
across the whole family.

By dissecting the discrepancy with a family of models\- that differ
only in their treatment of the induced Pt moment (Fe--Fe only, rigid
Pt moment, and the configuration-dependent Mryasov's and Polesya's
models\-), we traced its origin. It is not the variation of the
induced-moment magnitude per se: the Mryasov's and Polesya's models\-
capture that variation and still leave the interlayer ferromagnetism
marginally stable. The decisive quantity is the strength of the Pt-mediated
interaction. Reproducing the spin-spiral energetics requires an
amplitude $\lambda \approx 3.3$ relative to the bare estimate, and
the general bound $\lambda \leq 2$, valid for any induced-moment
model with an intra-atomic energy, shows that this enhancement
cannot originate from the Pt on-site physics. The energetics of the
induced moments are instead governed by the dressed
(exchange-enhanced) susceptibility, of which the frozen-potential
MFT captures only the bare part --- roughly one third. For
induced-moment magnets, the bare-versus-dressed distinction is
therefore not guaranteed to remain a quantitative correction: in
FePt, where the interlayer ordering rests entirely on the
Pt-mediated channel, it decides the magnetic ground state, while in
FePd, CoPt, and CoPd it shifts the ordering temperatures by several
hundred kelvin.

We have also discussed how the two parameter sets relate to the
established concepts of the magnetic force theorem, local
constraining fields, and static versus dynamical susceptibilities,
and we have presented Curie temperatures and magnon spectra for the
full L1$_0$ family as predictions for future experiments. The
applied theoretical methods are well established and applicable to
any magnetic material; the comparative analysis presented here ---
in particular the diagnostic amplitude $\lambda$ and its bound ---
may help the reader to choose, and to test, the appropriate approach
for other magnets with induced moments. This caution is particularly
relevant for simulations of complex noncollinear states such as
skyrmions, whose stability rests on delicate balances between
competing interactions: in materials with induced $4d/5d$ moments,
the contribution that the bare MFT parameters miss can be exactly
the one that tips such balances.

Our results can be important for a broader group of magnetic materials characterized by coexistence of well-defined local magnetic moments and strongly polarizable and more itinerant components. Other examples include 3$d$-metal oxides [\onlinecite{Logemann2017,Logemann2018}], as well as two-dimensional ferromagnet Fe$_3$GeTe$_2$ where exchange interaction between more localized and less localized magnetic moments at nonequivalent Fe atoms is relevant for stabilization of ferromagnetic ground state [\onlinecite{Katanin2026}]. This situation seems to be a very close analog of FePt case in our consideration.

\section{Method}
Throughout this work, the configuration-dependent exchange couplings are analyzed within a spin model that excludes spin--orbit coupling (SOC). The effect of SOC re-enters through the uniaxial magnetic anisotropy, which we add back when computing the magnon spectra and Curie temperatures. 
Although the magnetocrystalline anisotropy of L1$_0$ FePt is large, its energy scale is small compared to the exchange interactions, so it has only a minor effect on $T_\mathrm{C}$ and the magnon dispersion.

We write all spin Hamiltonians as sums over distinct unordered pairs
$\langle i,j\rangle$, each pair counted once and including every species
combination,
\begin{equation}
  H = -\sum_{\langle i,j\rangle} J_{ij}\,\hat{\mathbf{e}}_i\cdot\hat{\mathbf{e}}_j
    = -\frac{1}{2}\sum_{i\neq j} J_{ij}\,\hat{\mathbf{e}}_i\cdot\hat{\mathbf{e}}_j ,
  \label{eq:Hpairs}
\end{equation}
where the $\hat{\mathbf{e}}_i$ are unit vectors, the moment magnitudes are absorbed into $J_{ij}$, and self-interaction ($i\!=\!j$) is excluded by construction. This convention applies uniformly to like-species (Fe--Fe, Pt--Pt) and cross-species (Fe--Pt) pairs and removes the explicit double-counting factor from the Hamiltonian.
For a list of all conventions used in this work refer to Appendix~\ref{sec:all_conventions}.
The magnetic anisotropy term can be written as
\begin{equation}
   \sum\limits_{i} K_{\mathrm{U}} (\hat{\mathbf{e}}_i\cdot\hat{\mathbf{e}}_{\mathrm{U}})^2,
    \label{eq:magnetic_anisotropy}
\end{equation}
where $K_{\mathrm{U}}$ (a negative constant) is the uniaxial magnetocrystalline anisotropy strength and $\hat{\mathbf{e}}_{U}$ is a unit vector in the direction of the easy axis (coinciding with the $\mathbf{a}_3$ lattice vector, see definitions further below).

The L1$_0$ crystal structure is an ordered tetragonal derivative of the face-centered cubic (fcc) structure, belonging to the space group (P4/mmm) (No. 123). It consists of alternating atomic layers of two chemical species stacked along the crystallographic $c$-axis, resulting in a layered arrangement that lowers the cubic symmetry to tetragonal. The L1$_0$ structure is adopted by several technologically important intermetallic compounds, including FePt, CoPt, FePd, and Mn-based alloys, where the reduced symmetry often gives rise to pronounced magnetocrystalline anisotropy.
Below, we exemplify the method using FePt as the explicit example. For the other systems, Fe is replaced by Co and Pt by Pd, as needed.

Bravais lattice basis vectors $\mathbf{a}_d$, atomic basis, and reciprocal lattice basis vectors $\mathbf{b}_d$ used throughout are defined as follows, expressed in Cartesian coordinates:
\begin{tcolorbox}
\begin{equation}
\label{eq:bravais-lattice}
\begin{array}{@{}l@{\qquad}l@{}}
\begin{aligned}
\mathbf{a}_1 &= a\,(1,0,0),\\
\mathbf{a}_2 &= a\left(\tfrac12,\tfrac12,0\right),\\
\mathbf{a}_3 &= c\,(0,0,1),
\end{aligned}
&
\begin{aligned}
\mathbf{b}_1 &= \frac{2\pi}{a}(1,-1,0),\\
\mathbf{b}_2 &= \frac{2\pi}{a}(0,2,0),\\
\mathbf{b}_3 &= \frac{2\pi}{c}(0,0,1).
\end{aligned}
\end{array}
\end{equation}

\smallskip
Atomic basis: Fe/Co at $(0,0,0)$; Pd/Pt at
$\left(\tfrac12,0,\tfrac12\right)$.
\end{tcolorbox}
\noindent
With these definitions, $\mathbf{b}_1$, $\mathbf{b}_2$, and $\mathbf{b}_3$ point from the Brillouin-zone center ($\Gamma$) in the direction of the zone-boundary points M, X, and Z, respectively.

Values for $K_{\mathrm{U}}$ as well as the lattice parameters $a$ and $c$ are listed in Table~\ref{table:moments_TC}.

\subsection{Self-consistent spin spiral calculations}

We employ the generalized Bloch theorem,\cite{Sandratskii1998} as implemented in the Elk software
package\cite{elk} to compute self-consistent spin spirals from density functional theory (DFT). A 
spin spiral is specified by a propagation vector $\mathbf{q}$ and a cone angle
$\theta$. Labeling each atom by the Bravais lattice vector $\mathbf{R}$ of its
cell and the sublattice offset $\boldsymbol{\tau}_{\mu}$ ($\mu\in\{\mathrm{Fe},
\mathrm{Pt}\}$), the magnetization direction is
\begin{equation}
  \begin{split}
  \hat{\mathbf{e}}_{\mathbf{R}\mu}
    =\,& \sin\theta\,\cos\big(\mathbf{q}\cdot(\mathbf{R}+\boldsymbol{\tau}_{\mu})\big)\,
         \hat{\mathbf{u}}_{1} \\
       +\,& \sin\theta\,\sin\big(\mathbf{q}\cdot(\mathbf{R}+\boldsymbol{\tau}_{\mu})\big)\,
         \hat{\mathbf{u}}_{2} \\
       +\,& \cos\theta\,\hat{\boldsymbol{\eta}} ,
  \end{split}
  \label{eq:spiral}
\end{equation}
where $\hat{\boldsymbol{\eta}}$ is the cone axis and
the unit vectors $(\hat{\mathbf{u}}_{1},\hat{\mathbf{u}}_{2})$ span the plane perpendicular to the cone axis.
The axis $\hat{\boldsymbol{\eta}}$ is defined in spin space and is not
associated with any crystallographic direction, as long as spin-orbit coupling is not included.
The choice of
$\hat{\boldsymbol{\eta}}$ is therefore an arbitrary global convention. The
propagation vector $\mathbf{q}$, by contrast, is a reciprocal-space vector. The
moments thus precess on a cone about $\hat{\boldsymbol{\eta}}$, their phase
advancing by $\mathbf{q}\cdot\mathbf{R}$ from cell to cell. Such a configuration
is generally incommensurate, but the generalized Bloch theorem\cite{Sandratskii1998}
allows it to be treated within the chemical unit cell.
For each $(\mathbf{q},\theta)$ we converge the charge and spin
densities self-consistently, obtaining the total energy $E(\mathbf{q},\theta)$
together with the self-consistent atomic moment magnitude
$M_{\mathrm{Pt}}$, while the length of Fe moments $M_{\mathrm{Fe}}$ is fixed for each $\mathbf{q}$ at its self-consistent value for the ferromagnetic (FM) state. Depending on $\mathbf{q}$, the value of $M_{\mathrm{Pt}}$ can vary between zero and its maximum value, corresponding to the FM state.

We map the spin-spiral energy dispersion onto an effective Fe-only Hamiltonian, written as a sum over
distinct unordered Fe--Fe pairs,
\begin{equation}
  H = -\sum_{\langle i,j\rangle} J_{ij}\,
       \hat{\mathbf{e}}_i\cdot\hat{\mathbf{e}}_j .
       \label{Eq:Heisenberg}
\end{equation}
For the conical configuration of Eq.~(\ref{eq:spiral}), the products entering Eq.~(\ref{Eq:Heisenberg}) reduce to
\begin{equation}
  \hat{\mathbf{e}}_i\cdot\hat{\mathbf{e}}_j
  = \cos^2\theta + \sin^2\theta\,\cos(\mathbf{q}\cdot\mathbf{R}_{ij}),
\end{equation}
with $\mathbf{R}_{ij}$ the vector connecting the two Fe sites. 

Since the unit cell contains only one magnetic Fe site, the
effective Heisenberg coupling parameters follow by Fourier inversion
of the spin-spiral energy at fixed cone angle
$\theta$.\cite{Kubler2017} In the convention of
Eq.~(\ref{Eq:Heisenberg}) they read
\begin{equation}
    J_{ij} = -\frac{2\,\Omega}{\sin^2 \theta}
    \int\limits_{\mathrm{BZ}} \! \!\frac{d^3 q}{(2\pi)^3}\,
    E(\mathbf{q}) \, e^{-i \mathbf{q} \cdot \mathbf{R}_{ij} } ,
    \qquad \mathbf{R}_{ij} \neq 0 ,
    \label{eq:Jij_FT}
\end{equation}
where $\Omega$ is the unit-cell volume and the integration is taken
over the Brillouin zone (BZ). The $\mathbf{q}$-independent part of
$E(\mathbf{q})$ integrates to zero for $\mathbf{R}_{ij} \neq 0$, so
the pair couplings are unaffected by the choice of energy reference;
the factor of 2 reflects the unordered-pair convention of
Eq.~(\ref{Eq:Heisenberg}). The PBE approximation to the exchange-correlation functional\cite{PBE} and  a $(10\times 10\times 10)$ shifted $k$-mesh (up to 1000 $k$-points depending on the $q$-vector) were used in these calculations. A $(9\times 9\times 9)$  $q$-mesh was used and the integral over that mesh was computed using the fourth-order Simpson method.

We adopt $\theta=90^{\circ}$ for the main analysis. Such planar spin spiral maximizes the transverse modulation, $\sin^2\theta=1$, and hence the
$\mathbf{q}$-dependent energy. It is also the configuration that drives the induced
Pt moment through its full range of response between the ferromagnetic (FM) state
($\mathbf{q}= 0 $) and the maximally modulated zone-boundary states where the Pt moment goes to zero. The spin-spiral calculations reveal finite Pt (and Pd) moments whose magnitude varies with the chosen $\mathbf{q}$-vector. In this way, the effect of the induced moments on the total magnetic energy can be fully taken into account in the mapping to the bilinear Heisenberg atomistic spin Hamiltonian. 

At the other extreme of possible choices of spiral cone angles, i.e., as $\theta \rightarrow 0$, there is a conceptual link to the magnetic-force-theorem approach. In the limit of an infinitesimal spin-spiral cone angle, both the magnetic-force-theorem and self-consistent spin-spiral approaches probe the quadratic response of the magnetic energy to transverse spin rotations. The essential difference is the electronic response that accompanies it. In the magnetic-force-theorem approach the effective potential is kept frozen, so the energy curvature is governed by the bare susceptibility. In contrast, the self-consistent spin-spiral calculation allows the electronic density --- and in FePt the induced Pt moments --- to relax self-consistently. The resulting curvature is therefore governed by the exchange-enhanced (dressed) susceptibility. Self-consistent spin-spiral calculations for additional cone angles are
collected in Appendix~\,\ref{sec:cone_angle-magnons}.

\subsection{Magnetic force theorem-based calculations of exchange coupling parameters}

The magnetic force theorem (MFT)\cite{Weinert1985,LKAG1987} forms the basis for a computationally efficient
and widely used approach for computing exchange couplings. This approach is vastly less computationally demanding compared to the spin-spiral approach described above. The MFT states that the
change in the total energy produced by an infinitesimal rotation of the magnetic
moments is, to first order, equal to the corresponding change in the
single-particle (Kohn--Sham) band energy evaluated with the effective potential
frozen at its self-consistent value in the reference state. This is a consequence
of the variational character of the density-functional total energy:
to first order in the perturbation, the explicit change of the potential and the
double-counting terms 
cancel, leaving only the change in the sum of occupied
one-electron energies. In practice it means that the response of the system to a
small spin tilt is captured by the rearrangement of the occupied states alone, so
that all pairwise couplings can be extracted from a single self-consistent
calculation of the reference state followed by a Green-function post-processing
step, without reconverging the charge and spin densities for every perturbed
configuration.
Thus, the exchange couplings $J_{ij}$ for any two sites are obtained from
second-order response theory in small spin deviations, as discussed in
[\onlinecite{LKAG1987,Katsnelson2000,Szilva2023}].
We employ MFT using the LKAG approach,\cite{LKAG1987} as implemented in the software package RSPt
[\onlinecite{Wills1987},\onlinecite{Wills2010}], an all-electron full-potential linear
muffin-tin orbital (FP-LMTO) method, to compute Heisenberg interaction parameters
$J_{ij}$ around the ferromagnetic reference state. 
The $J_{ij}$ are obtained using the expression
\begin{equation} J_{ij}=\frac{T}{4} \sum_{n} \operatorname{Tr}\left[\hat{\Delta}_{i}\!\left(i \omega_{n}\right)\hat{G}_{ij}^{\uparrow}\!\left(i \omega_{n}\right) \hat{\Delta}_{j}\!\left(i \omega_{n}\right) \hat{G}_{ji}^{\downarrow}\!\left(i \omega_{n}\right)\right]. \label{eqn2} \end{equation}
Here, $T$ denotes the temperature, $\hat{\Delta}$ is the on-site exchange potential describing exchange splitting at sites $i$ and $j$, and $\hat{G}_{ij}^{\sigma}$ is the intersite Green’s function for spin $\sigma$ (either $\uparrow$ or $\downarrow$). The $n^\mathrm{th}$ fermionic Matsubara frequency is represented by $\omega_n$. These terms are matrix elements in both orbital and spin space, with the trace taken over orbital indices. The temperature $T$ entering Eq.~(\ref{eqn2}) acts as a smearing parameter (set at 1~mRy in our calculations) that sets the spacing of the Matsubara frequencies and thereby the number of frequencies required for convergence. A lower smearing temperature narrows the energy window and demands a higher frequency cutoff, whereas a larger value converges faster at the expense of a possible smearing dependence of the computed $J_{ij}$.
We emphasize that the electronic structure
quantities here (intersite Green's functions $\hat{G}_{ij}$ and on-site spin splittings
$\hat{\Delta}_i$) are calculated assuming a specific reference state, in this case
ferromagnetic (FM), and, in general, depend on that reference state
[\onlinecite{Szilva2023,Logemann2017,Logemann2018}].

This reference-state dependence is central to the present work. Because the Pt
moment is itself induced by the surrounding Fe order, couplings evaluated at the FM
reference are obviously not guaranteed to describe configurations far from it. This motivates
both the comparison with the self-consistent spin-spiral results, in which the
induced Pt moments are allowed to relax, and the configuration-dependent models introduced
in previous sections.

In principle, the reference state does not need to be the
ferromagnetic one. The couplings can equally be computed around an
antiferromagnetic (AFM), a disordered-local-moment (DLM), or even a spin-spiral
configuration. Each choice freezes the electronic structure --- and hence all moment
magnitudes, including the induced Pt moment --- at that state's self-consistent
values, yielding a distinct, reference-dependent set of $J_{ij}$.

The approach can also be combined with constrained or fixed-spin-moment schemes.
Using constrained DFT or a penalty field, the moment magnitude can be held at a
prescribed, non-equilibrium value, with the $J_{ij}$ evaluated around it, rather than
around the self-consistent magnitude of the reference state.

By construction, the magnetic
moment magnitude is fixed at its value for the ferromagnetic reference state, so
that the $J_{ij}$ describe the energetics of transverse rotations of rigid moments
about the collinear ferromagnetic (FM) configuration.  Specifically, it is important to note that in this method the induced Pt moment is held
at its FM value and is not allowed to relax as the Fe spins are tilted. Thus,
longitudinal fluctuations of the induced Pt moment are not addressed with this
method.
In the present computations, exchange and
correlation are treated within the PBE  approximation for the exchange-correlation functional
\cite{PBE}, and the Brillouin-zone integration employs a shifted
$(40\times 40\times 20)$ $k$-mesh (32{,}000 points, reduced to 16{,}000 in the
irreducible Brillouin zone).

\subsection{Critical temperature}
The Curie temperature was determined from classical Monte Carlo (Metropolis) and ASD simulations of the spin Hamiltonian in Eq.~(\ref{eq:Hpairs}), supplemented by the uniaxial anisotropy term in Eq.~(\ref{eq:magnetic_anisotropy}), as implemented in the UppASD atomistic spin-dynamics software package.\cite{eriksson2017atomistic} The temperature dependence of the magnetization, $M(T)$, and the magnetic susceptibility were calculated over a range of temperatures. The Curie transition is visually reflected in the steepest decrease of $M(T)$ and a corresponding peak in the susceptibility. The reported $T_\mathrm{C}$ values are obtained from a finite-size scaling analysis using the fourth-order Binder cumulant,\cite{binder2012monte} which provides a numerically accurate estimate of the critical temperature.

\subsection{Magnon spectra}

The dynamical structure factor, $S^k(\mathbf{q},\omega)$, is the central quantity describing the magnetic excitation spectrum. It is directly accessible in inelastic neutron-scattering experiments and, for thin films, by spin-polarized electron energy-loss spectroscopy (SPEELS) \cite{eriksson2017atomistic,etz2015atomistic}. It therefore provides a direct link between the calculated spin dynamics and experiment.
The space- and time-resolved spin correlation function is defined as
\begin{equation}
C^{k}(\mathbf{r}-\mathbf{r}',t) =
\langle m^k_{\mathbf{r}}(t)\,m^k_{\mathbf{r}'}(0)\rangle
-
\langle m^k_{\mathbf{r}}(t)\rangle
\langle m^k_{\mathbf{r}'}(0)\rangle,
\end{equation}
where $\langle\cdots\rangle$ denotes an ensemble average and $k$ a Cartesian component. The dynamical structure factor is obtained as the space-time Fourier transform of the correlation function,
\begin{equation}
S^k(\mathbf{q},\omega)=
\frac{1}{\sqrt{2\pi}\,N}
\sum_{\mathbf{r},\mathbf{r}'}
e^{i\mathbf{q}\cdot(\mathbf{r}-\mathbf{r}')}
\int\limits_{-\infty}^{\infty}
e^{i\omega t}
C^{k}(\mathbf{r}-\mathbf{r}',t)\,dt.
\end{equation}

The correlation functions are evaluated from finite-temperature atomistic spin-dynamics simulations based on the stochastic Landau--Lifshitz--Gilbert equation, as implemented in the UppASD package\cite{eriksson2017atomistic}, using spin Hamiltonian parameters obtained from the spin-spiral calculations as well as from the MFT calculations. The corresponding magnon dispersion is extracted from the energies at which $S^k(\mathbf{q},\omega)$ exhibits maximum intensity for a given wave vector $\mathbf{q}$.

Experimental information on magnon dispersions is presently very limited for the L1$_0$ alloys considered here, despite their considerable technological and fundamental importance. The calculated dynamical structure factors therefore provide predictions that can serve as benchmarks for future neutron-scattering or SPEELS measurements.

For comparison, we also calculate the adiabatic magnon spectra within linear spin-wave theory \cite{eriksson2017atomistic,etz2015atomistic}, which is expected to be accurate in the low-temperature limit and for weak Gilbert damping.

\subsection{Spin stiffness}

We calculate the spin stiffness using two methods:
\begin{enumerate}
    \item from the curvature of the spin-spiral energy as a function of spiral wavevector $\vec{q}$ calculated using Elk code along the [100] direction.
    \item by direct summation of interatomic exchange interactions $A = \frac12 \lim\limits_{\mu\to 0} \sum\limits_{\langle i,j \rangle} J_{ij}R_{ij}^2 \exp(-\mu R_{ij})$, where $R_{ij}$ is the distance between interacting spins and $\mu$ is the parameter of the exponential function that makes the sum over neighboring spins converge (both quantities $J_{ij}$ and $R_{ij}$ are in Eqns.~\ref{Eq:Heisenberg} and \ref{eq:Jij_FT}). Without the regularizing exponential function, the direct summation diverges upon inclusion of more interacting shells (see e.g.~discussion in [\onlinecite{Pajda2001},\onlinecite{Borisov2024rev}]).

    The $\mu\to 0$ limit of the spin stiffness depends, in principle, on the range of $\mu$ values that were used for extrapolation towards $\mu = 0$, so in Table~\ref{table:stiffness} we report the results corresponding to the range $2 \leq \mu \leq 3$ and including also the induced Pt or Pd rigid moments in the Heisenberg model. Variations around 10--15\% depending on the choice of initial $\mu$ values can be observed in that procedure but the general trends among the 4 studied systems should hold.
\end{enumerate}

\section*{Acknowledgements}

Financial support from the Swedish Research Council (Vetenskapsrådet, VR; Grants No. 2024-04986 and No. 2024.05206), Carl Tryggers Foundation (grant number CTS 22:2013), the Knut and Alice Wallenberg Foundation (KAW; Grant No. 2022.0108), and Olle Engkvist Foundation are acknowledged. 
The Wallenberg Initiative Materials Science for Sustainability (WISE) funded by the Knut and Alice Wallenberg Foundation and The Swedish e-Science Research Centre (SeRC) are acknowledged too.

The computations/data handling were enabled by resources provided by the National Academic Infrastructure for Supercomputing in Sweden (NAISS), partially funded by the Swedish Research Council through grant agreement no. 2022-06725.

Discussions with Lars Nordström, Danny Thonig, Zhiwei Lu, and Ivan Miranda are acknowledged.

\section*{Data availability}
The exchange parameter sets presented in this study are available through \url{mintatlas.org}.

\appendix

\section{$\mathrm{Fe}/\mathrm{Co}-\mathrm{Pd}/\mathrm{Pt}$ and $\mathrm{Pd}/\mathrm{Pt}-\mathrm{Pd}/\mathrm{Pt}$ Heisenberg exchange parameters}
\label{sec:further_interactions}

\begin{figure}[h]
\vspace{10pt}
{\centering
\includegraphics[width=0.46\textwidth]{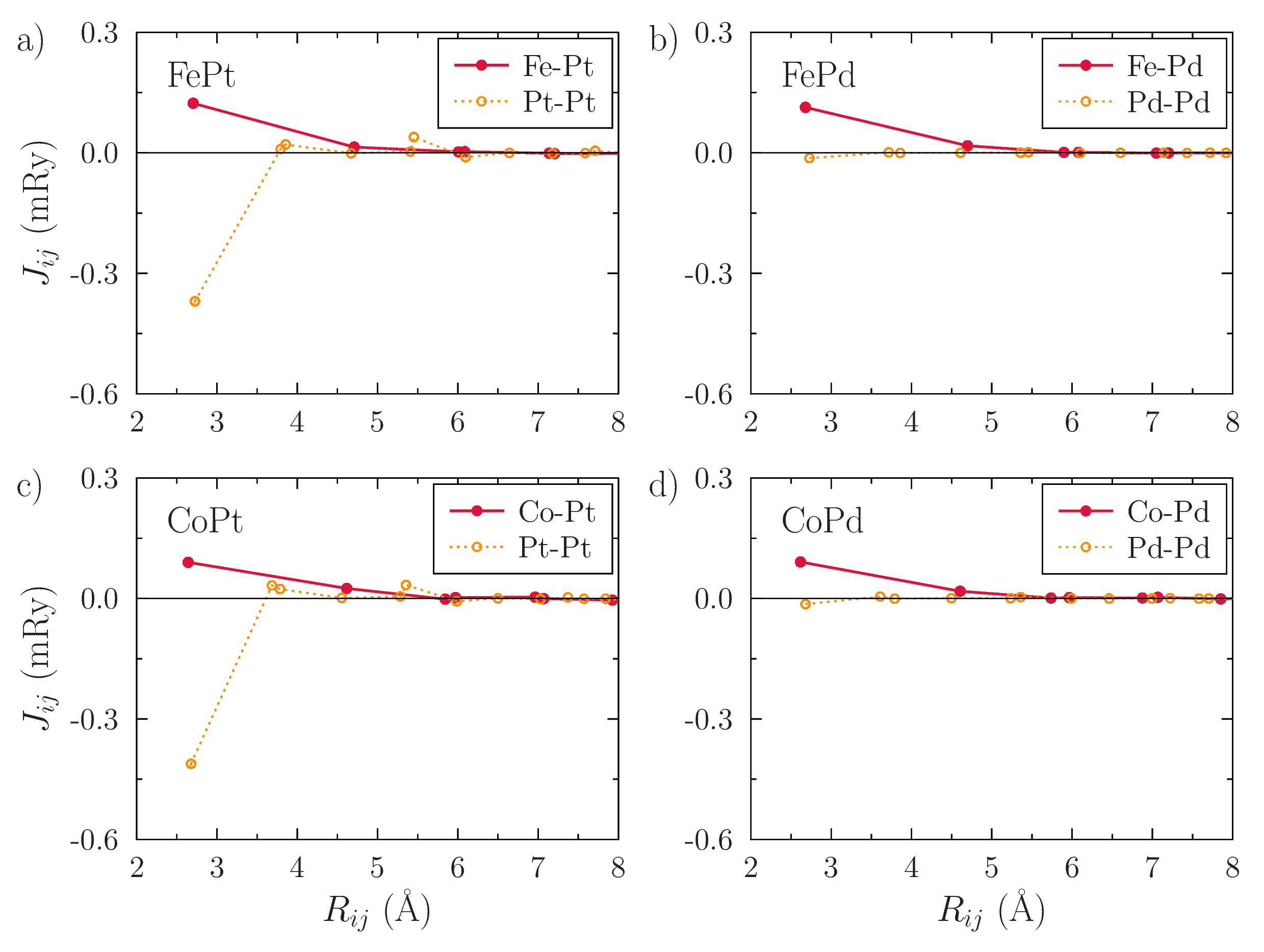}
}
\vspace{-5pt}
\caption{
Heisenberg interactions between $3d$ transition metal spins and $4d/5d$ induced moments as well as interactions among the induced moments in the studied L1$_0$ compounds calculated using the magnetic force theorem (MFT) in RSPt code.}
\label{f:all_Jijs}
\end{figure}

Fig.\,\ref{f:all_Jijs} shows Fe-Pt and Pt-Pt Heisenberg exchange parameters obtained from magnetic force theorem, which complement the data shown in Fig.~\ref{fig:Jij_all}. The Fe-Pt interactions are mostly ferromagnetic, while the nearest-neighbor Pt-Pt interaction is antiferromagnetic. Same trend is observed for CoPt, while Pd-Pd interactions are very weak in FePd and CoPd.

\section{Isotropic part of the energy of the delocalized $\mathrm{Pt}$ states}
\begin{figure}[H]
\includegraphics[width=0.43\textwidth]{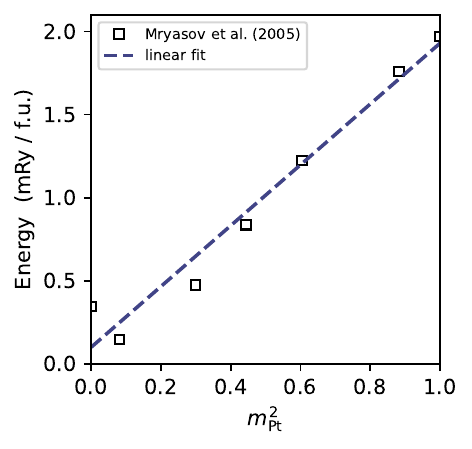}
\caption{Isotropic part of the energy of the delocalized Pt states
in FePt as a function of the squared reduced Pt moment
$m_{\mathrm{Pt}}^{2}$, digitized from Fig.~1(b) of
Ref.~\onlinecite{Mryasov2005} (squares), together with a
least-squares linear fit (dashed line). Energies in mRy per formula
unit.}
\label{fig:mry_fig1b}
\end{figure}

\section{$\mathrm{FePt}$ model with rigid $\mathrm{Pt}$ moments: spin-spiral energy expression for an arbitrary cone angle}
\label{sec:cone_angle_formulas}

For a conical spiral the Fe moments make a common cone angle $\theta$ with the
spiral axis, $\hat{\mathbf{e}}_{i} = (\sin\theta\cos(\mathbf{q}\cdot\mathbf{R}_{i}),\,
\sin\theta\sin(\mathbf{q}\cdot\mathbf{R}_{i}),\,\cos\theta)$, with $\theta=90^{\circ}$
recovering the flat spiral. The rigid Pt moment is taken to point along its
instantaneous exchange field from the surrounding Fe (the Pt--Pt couplings are evaluated at this slaved configuration
but do not act back on the Pt direction). The field has a uniform component
$\propto\cos\theta$ along the axis and a rotating transverse component
$\propto\sin\theta\, \bar{h}(\mathbf{q})$, with $\bar{h}(\mathbf{q})$
the normalized exchange-field magnitude of Eq.~(\ref{eq:mry_hbar}),
so that the Pt sublattice itself forms a cone with the reduced angle
$\alpha(\mathbf{q},\theta)$ given by
$\tan\alpha = \bar{h}(\mathbf{q})\tan\theta$.
The per-formula-unit energy relative to the ferromagnetic state becomes
\cite{Sandratskii1998,lyons1960}
\begin{equation}
  \begin{split}
  \Delta E(\mathbf{q},\theta)
   ={}& \sin^{2}\!\theta\,\cdot \tfrac{1}{2}\sum_{j} J^{\mathrm{FeFe}}_{0j}
        \big[1-\cos(\mathbf{q}\cdot\mathbf{R}_{j})\big] \\
      +&\; J_{\mathrm{FePt}}
        \Big[1-\sqrt{\cos^{2}\!\theta+\sin^{2}\!\theta\; \bar{h}^{2}(\mathbf{q})}\,\Big] \\
      +&\; \sin^{2}\!\alpha(\mathbf{q},\theta)\,\cdot \tfrac{1}{2}\sum_{j}
        J^{\mathrm{PtPt}}_{0j}
        \big[1-\cos(\mathbf{q}\cdot\mathbf{R}_{j})\big],
  \end{split}
  \label{eq:dE_cone}
\end{equation}
with
$\sin^{2}\!\alpha = \sin^{2}\!\theta\, \bar{h}^{2}/
\big(\cos^{2}\!\theta+\sin^{2}\!\theta\, \bar{h}^{2}\big)$.
The first term is the $\sin^{2}\theta$ scaling
of a bilinear interaction on a common cone. The second is the energy of the
rigid Pt moment in a field of relative magnitude
$\sqrt{\cos^{2}\theta+\sin^{2}\theta\, \bar{h}^{2}}$. The third is the Pt--Pt
contribution evaluated on the reduced Pt cone $\alpha\le\theta$.
At $\theta=90^{\circ}$ it
reduces to Eq.~(\ref{eq:dE_models}) wherever the exchange-field sum
inside the absolute value in Eq.~(\ref{eq:mry_hbar}) is non-negative.
In the opposite, small-cone limit,
\begin{equation}
\begin{split}
\lim_{\theta\to 0}
\frac{\Delta E(\mathbf{q},\theta)}{\sin^{2}\theta}
={}&
\frac{1}{2}
\sum_{j} J^{\mathrm{FeFe}}_{0j}
\bigl[1-\cos(\mathbf{q}\cdot\mathbf{R}_{j})\bigr]
\\
&+
\frac{1}{2}J_{\mathrm{FePt}}
\bigl[1-\bar{h}^{2}(\mathbf{q})\bigr]
\\
&+
\frac{\bar{h}^{2}(\mathbf{q})}{2}
\sum_{j} J^{\mathrm{PtPt}}_{0j}
\\
&\qquad\times
\bigl[1-\cos(\mathbf{q}\cdot\mathbf{R}_{j})\bigr].
\end{split}
\label{eq:dE_cone_limit}
\end{equation}
which, up to the small Pt--Pt correction, is the Mryasov's model with the non-enhanced amplitude $\lambda=1$.

\section{Mryasov model with optimized $\lambda$ fit in each direction}
\label{sec:Mryasov_model}

\begin{figure}[h]
{\centering
\includegraphics[width=0.39\textwidth]{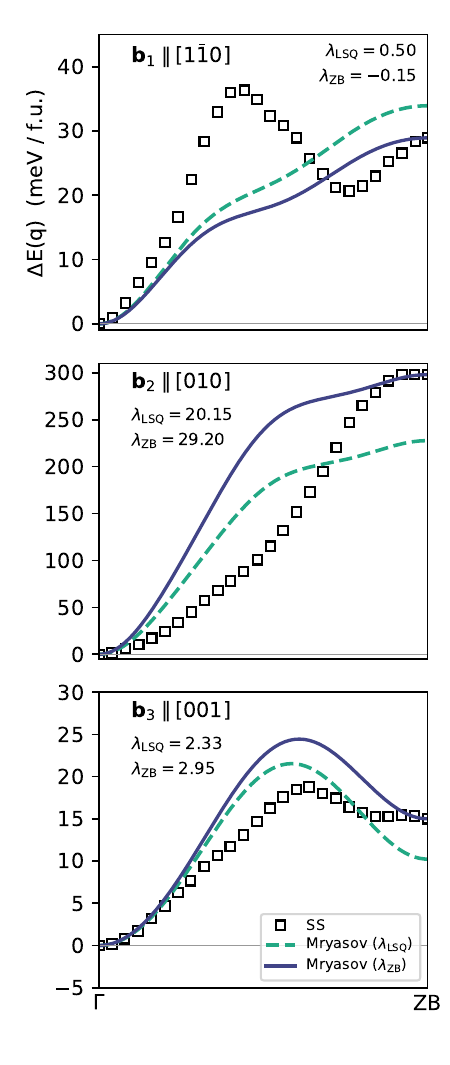}
}
%\vspace{-20pt}
\caption{Fits of the Mryasov model to the spin spiral data providing values for $\lambda$ per direction. LSQ = a least squares fit to all points. ZB = the zone boundary spin spiral energy value is enforced.
}
\label{fig:mryasov_lambdafit}
\end{figure}

Fig.\,\ref{fig:mryasov_lambdafit} shows the Mryasov model with optimized/fitted lambda in each direction.
The Mryasov model is built to correct the $\mathbf{b}_3$ direction. The lambda-values obtained by fitting the other direction are very different, signaling that the Mryasov model is not a general model meant to mimic the spin-spiral results in every direction.
%\AnnaD{There is more to say here.}

\section{MFT dataset convergence tests}
\label{si:mft_convergence}

\begin{figure}[h]
{\centering
\includegraphics[width=0.39\textwidth]{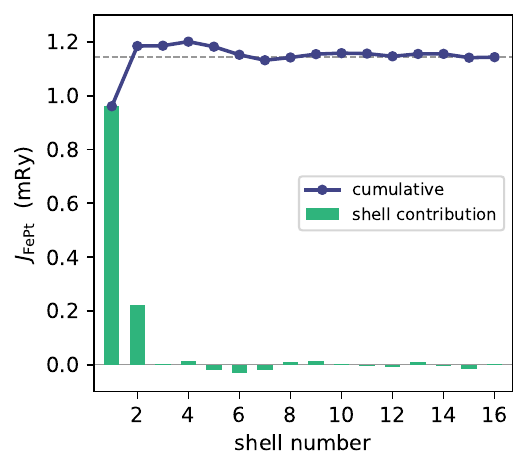}
}
%\vspace{-20pt}
\caption{ Convergence of $J_{\mathrm{FePt}}$ with respect to the number of shells. $J_{\mathrm{FePt}}$ converges to 1.143\,mRy.
}
\label{fig:jfept_convergence}
\end{figure}

Fig.\,\ref{fig:jfept_convergence} shows convergence of $J_{\mathrm{FePt}}$ as a function of shell number. We see that the first two shells are the main contributors and that the sum appears converged to within a few percent at 16 shells.

\begin{figure}[h]
{\centering
\includegraphics[width=0.39\textwidth]{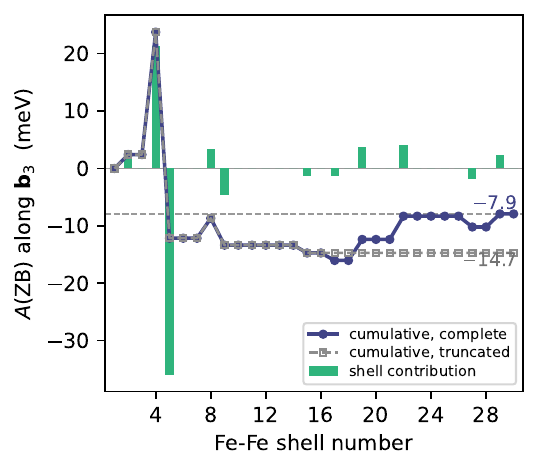}
}
%\vspace{-20pt}
\caption{ 
Convergence of the MFT  Fe-Fe $J_{ij}$ data set.
$A(\mathrm{ZB}) = 1/2 \sum_j J^{\mathrm{FeFe}}_j [1 - \cos{(
\mathbf{q}_{\mathrm{ZB}} \cdot \mathbf{R}_j)
}]$ is plotted with respect to number of shells included. The zone boundary 
$\mathbf{q}$
in the $\mathbf{b}_3$ direction is used to construct this test of the robustness of the AFM ground state for this spin Hamiltonian. 
}
\label{fig:fig_azb_b3_convergence}
\end{figure}

Fig.\,\ref{fig:fig_azb_b3_convergence} show the Fe-Fe interaction convergence as a function of shell number. Clearly, the sum converges very slowly and an error at the level of a few meV is possible, even if interaction shels up to 3 lattice constants are included (dark purple dots).  The conclusion however is that the negative sign of $A(\mathrm{ZB})$ along
$\mathbf{b}_3$ would likely prevail even if additional shells were included. Truncating the dataset about half-way out shell-wise (gray squares) provides a more strongly AFM state compared to including all interactions up to 3 lattice parameters (30 shells in Fig.~\ref{fig:fig_azb_b3_convergence}). The fifth shell is the main contributor of AFM coupling, but several more distant shells contribute to this too.

\section{Spin spiral convergence}
\label{sec:ss_convergence}

\begin{figure}[h]
\includegraphics[width=0.43\textwidth]{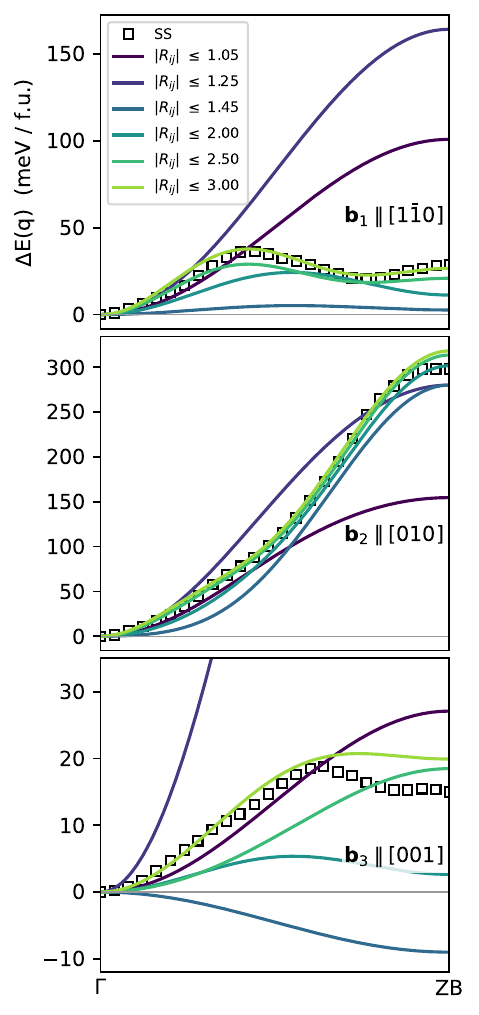}
\caption{
Convergence test of the spin-spiral calculations along $\mathbf{b}_1$, $\mathbf{b}_2$, and $\mathbf{b}_3$. The legend gives the real-space cutoff radius $|R_{ij}|$ of the Fe--Fe interaction sum, in units of the lattice parameter $a$. Summation up to $3a$ is clearly needed, corresponding to over 200 Fe-Fe interactions for each Fe atom. There is in fact still room for improvement, despite the numerous included interactions. This is especially visible along the $\mathbf{b}_3$ direction.
 }
\label{fig:shell_cutoff_test}
\end{figure}
Here, we show how the spin spiral energy converges as an increasing number of interaction shells are included. In summary, shells up to at least $3a$ are needed, which corresponds to more than 200 Fe neighbors (30 shells) with which each Fe atom interacts. Along the $\mathbf{b}_3$ direction (lowest panel in Fig.\,\ref{fig:shell_cutoff_test}), there is actually still room for improvement close to the zone boundary, i.e., including even more shells would be beneficial. The very large number of interactions needed to achieve convergence is relevant to keep in mind when constructing effective spin Hamiltonians using methods such as energy mapping or the four-state approach.\cite{Li2021SpinHamiltonians}

\section{Effect of changing the cone angle}
\label{sec:cone_angle-magnons}

\begin{figure}[h]
\vspace{-10pt}
\includegraphics[width=0.45\textwidth]{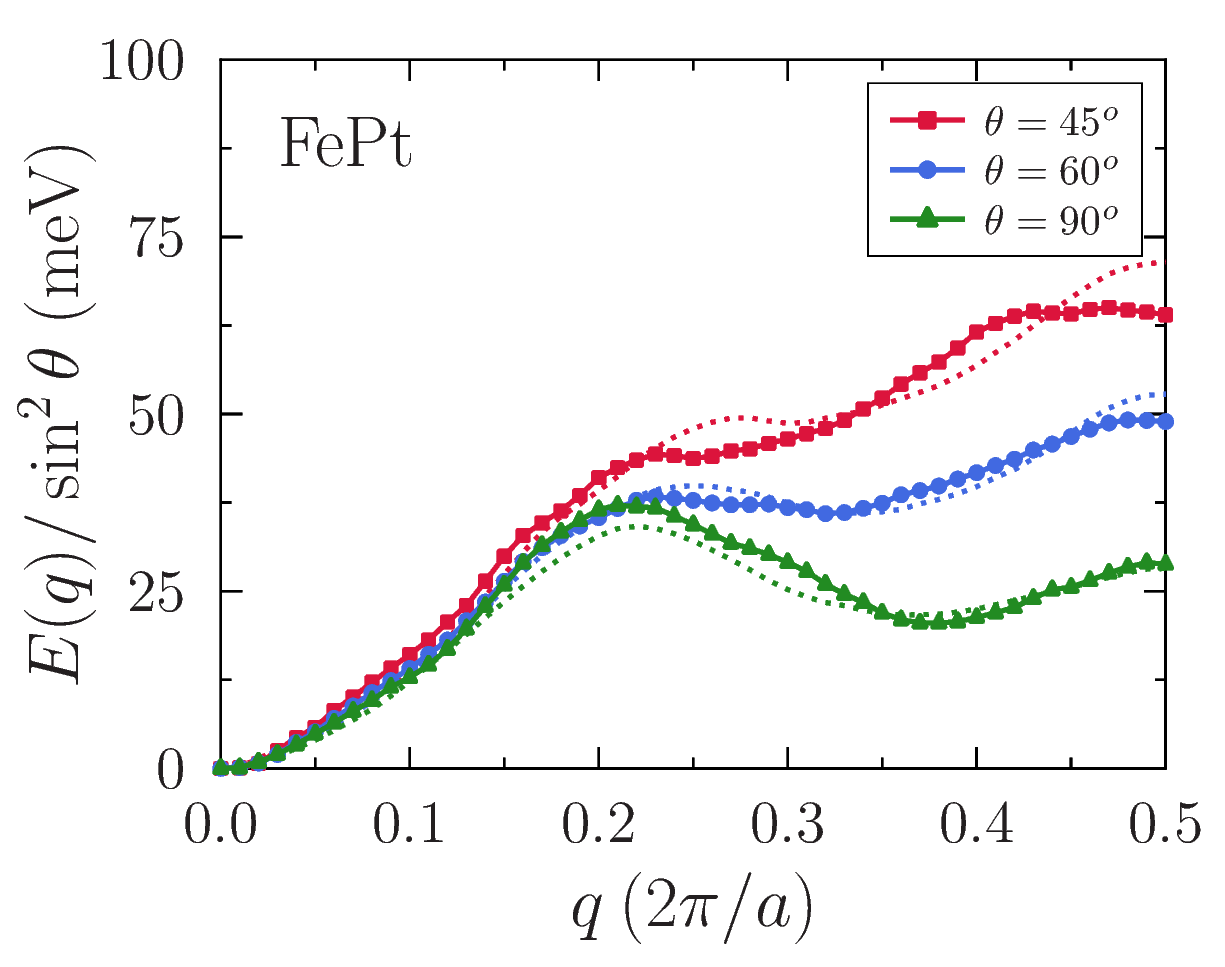}
\vspace{-10pt}
\caption{
Spin-spiral energies (divided by $\sin^2\theta$) along the [100] direction calculated using Elk code for FePt and different cone angles $\theta$. The length of the Fe moments was constrained to the equilibrium value calculated for the ferromagnetic state. Solid lines correspond to $(10\times 10\times 10)$ $k$-mesh and dashed lines~-- to $(20\times 20\times 20)$ $k$-mesh, demonstrating well converged results.
}
\label{fig:spirals_different_theta}
\end{figure}

In this section, the spin-spiral energies for FePt and different angles $\theta$ between atomic moments and the $z$-axis are presented. While we focus on $\theta = \pi/2$ in the main text, we also did calculations for further values of $\theta$ (45\textdegree{} and 60\textdegree{}) and the results are plotted in Fig.~\ref{fig:spirals_different_theta}.

\begin{figure}[h]
\includegraphics[width=0.45\textwidth]{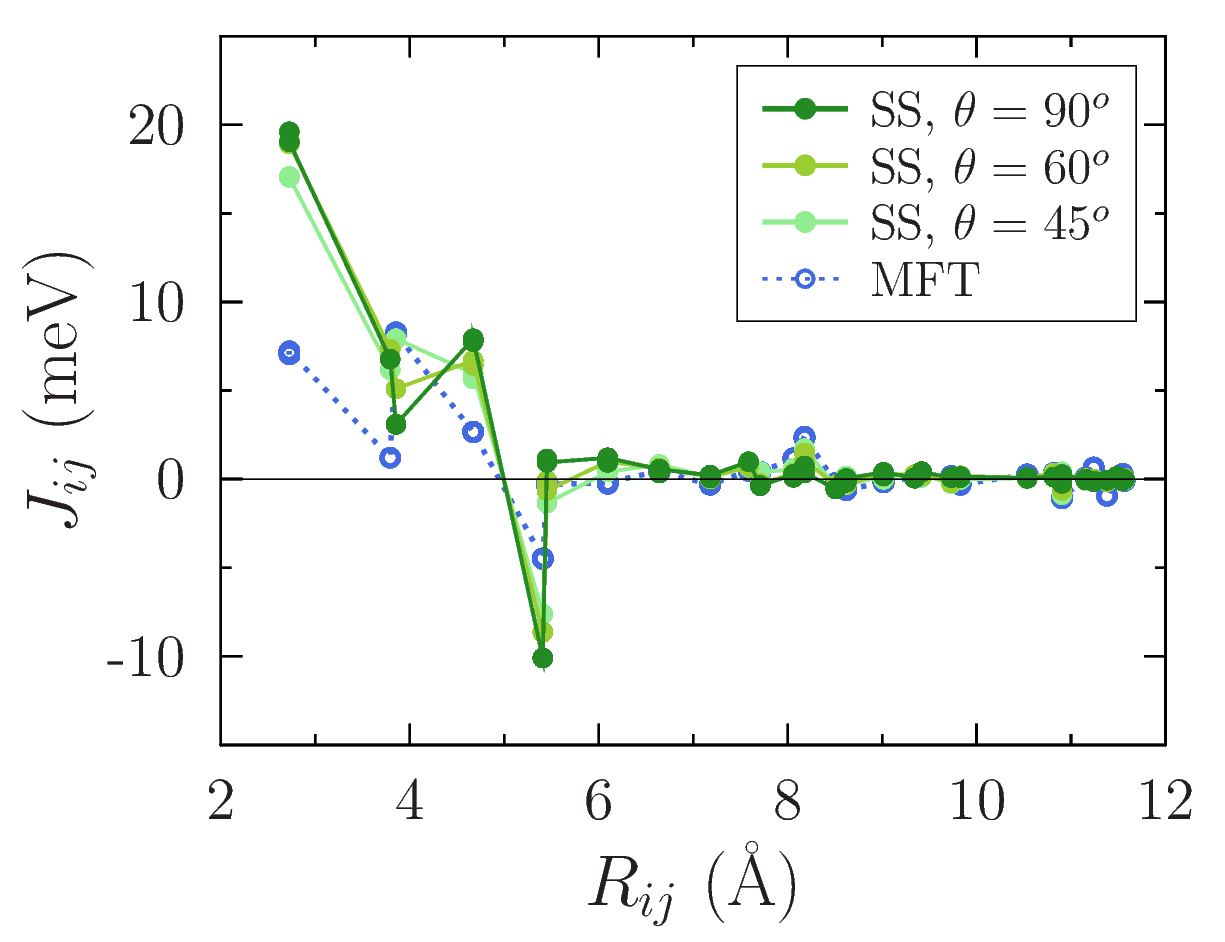}
\vspace{-10pt}
\caption{
Effective Heisenberg Fe-Fe interactions for FePt from spin spirals with different cone angle $\theta$, in comparison with values calculated using magnetic force theorem (MFT).}
\label{fig:Jij_different_theta}
\vspace{-10pt}
\end{figure}

The key difference to the $\theta=\pi/2$ results is that the Fe moments around each Pt are never fully compensated, meaning that the induced Pt moment does not reach zero for angles $\theta < \pi/2$. Accordingly, the $\vec{q}$-dependence of the Pt moment is weaker for lower the values of the angle, $\theta$. This turns out to have a marked effect on the spin-spiral energies (Fig.~\ref{fig:spirals_different_theta}), demonstrating important contribution of Pt moments to the magnetic energy and effective Fe-Fe interactions especially for excitations close to the zone-boundary. Close to the zone-center the different values of $\theta$ give rather similar results, but as we will see below the difference is sufficiently large to result in quite different values of the interatomic exchange interaction. To illustrate this we show in  Fig.~\ref{fig:Jij_different_theta} these effective Heisenberg interactions, plotted vs distance for different angles $\theta$ in the calculations.

\begin{figure}[H]
\includegraphics[width=0.45\textwidth]{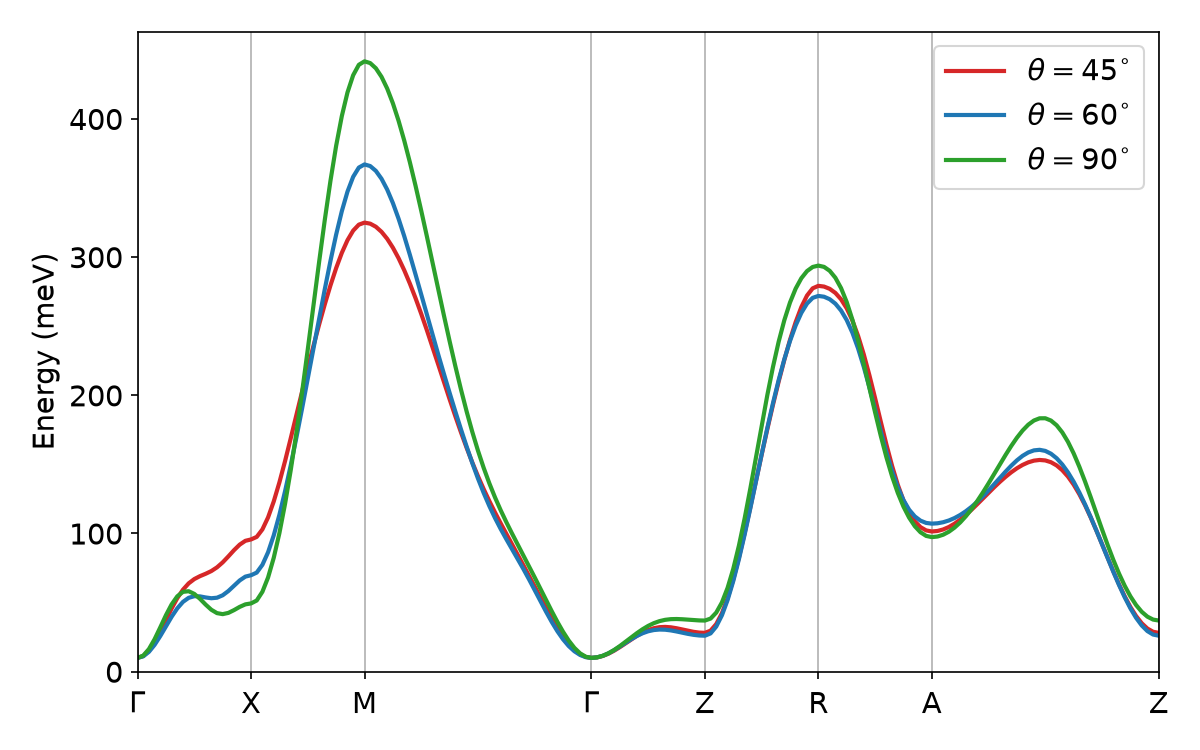}
\caption{Adiabatic magnon spectra  calculated at $T=0\,\mbox{K}$ for FePt from spin spirals with different angles $\theta$ specified in the figure.}
\label{fig:ams_different_theta}
\end{figure}

These results demonstrate that the renormalization of effective interactions due to induced Pt moments is slightly reduced for smaller angles $\theta$, which is to be expected due to smaller variations of those induced moments. It should be pointed out that calculations for even smaller angles of $\theta$; 5\textdegree{} and 10\textdegree{}, revealed numerical instabilities most likely due to the fact that the spin-spiral energies are much smaller and putting impractically high demands on convergence criteria. For this reason, these results are not shown. In addition, in Fig.~\ref{fig:ams_different_theta} we demonstrate the impact of the $\theta$ values on the adiabatic magnon spectra. Finally, the interactions with $\theta = 90^{\circ}$ results in a slightly lower $T_\mathrm{C}$ in comparison to $\theta = 45^{\circ}$.

\newpage
\section{Effect of lattice parameter and basis set}
\label{sec:emto}

Here, we show a comparison of the Heisenberg Fe-Fe interactions coupling for the FM L1$_0$-ordered structure of FePt, considering different lattice parameters and DFT methods. The exact muffin-tin orbital (EMTO) code has been used to calculate the magnetic exchange interactions  
at 0 K within the magnetic force theorem \cite{Liechtenstein1984,LKAG1987}, as implemented in the Lyngby version \cite{Ruban2016}. The one-electron Kohn-Sham equations were 
solved within the soft-core and scalar-relativistic approximations, with $l_{\rm max} = 3$ for partial waves
and $l_{\rm max}^{\rm t} = 5$ for their ``tails''. The Green's function was calculated for 16 complex energy points distributed exponentially on a semi-circular contour including states within 1~Ry below the Fermi level.
The exchange-correlation effects were described within the local spin-density approximation \cite{LDA1, LDA2}.
\begin{figure}[H]
\includegraphics[width=0.45\textwidth]{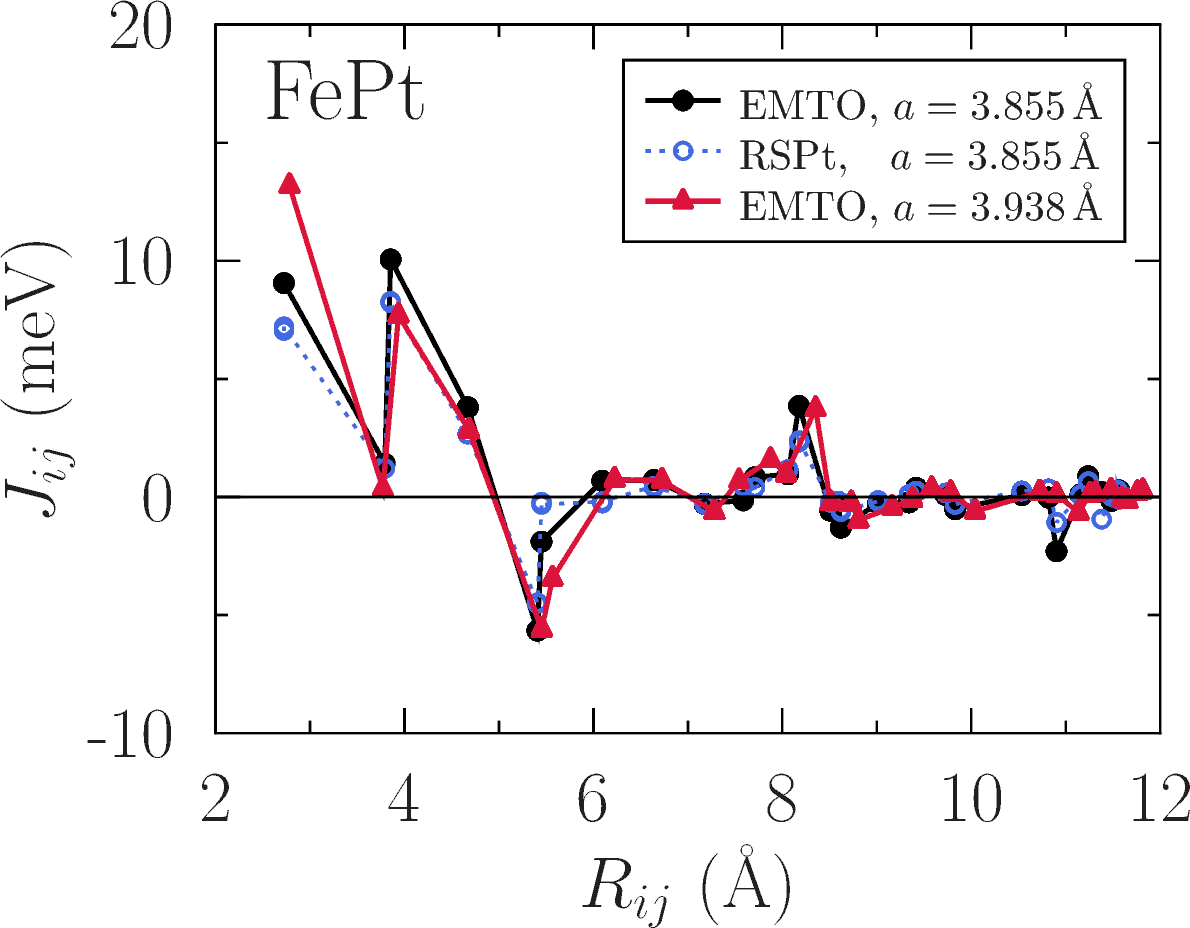}
\caption{
Heisenberg Fe-Fe interaction for L1$_0$-ordered alloy FePt computed with EMTO and RSPt codes. Different lattice parameters were considered, where the data for $a$ = 3.855 Å and $c/a = 0.984$ (experimental values) are plotted in black and blue, and the data for $a$ = 3.938 Å and $c/a=0.958$ (optimized EMTO values) are plotted in red.}
\label{fig:compare_emto_rspt}
\end{figure}

The Fe-Fe exchange coupling is compared in Fig.~\ref{fig:compare_emto_rspt} for both EMTO and RSPt calculations. Here, the same lattice parameter used in previous RSPt calculations (a$_{\mathrm{RSPt}}$ = 3.855 Å) was considered, along with the optimized value obtained by fitting the Morse equation of states \cite{Morse} in EMTO calculations (a$_{\mathrm{EMTO}}$ = 3.938 Å). Interestingly, the J$_{ij}$ nearest-neighbor interaction increases by 45.5\% when changing the lattice parameter from a$_{\mathrm{RSPt}}$ to a$_{\mathrm{EMTO}}$, while EMTO and RSPt results are similar if the same lattice parameter is used. This illustrates that two different methods give the same results, and that the failure of the LKAG approach in reproducing experimental observation of FePt is not due to an error in a specific software implementation. The results of Fig.~\ref{fig:compare_emto_rspt} also illustrate a strong exchange striction of the FePt system.

\section{Conventions used}
\label{sec:all_conventions}

Table~\ref{tab:conventions} collects the notation and conventions
used throughout, together with conversion rules to other common
conventions.

\newpage
\renewcommand{\arraystretch}{1.15}
\setlength{\tabcolsep}{9pt}
\begin{table*}
\caption{Notation and conventions used throughout this work. The
last column gives units, defining equations, and conversion rules to
other common conventions.}
\label{tab:conventions}
\centering
\begin{tabular}{p{0.16\textwidth}p{0.36\textwidth}p{0.42\textwidth}}
\hline
Symbol & Meaning & Convention / units / notes \\
\hline
$\hat{\mathbf{e}}_i$ & direction of the Fe (Co) moment at site $i$ &
unit vector; moment magnitudes are absorbed into the couplings \\[5pt]
$\langle i,j\rangle$ & sum over distinct unordered pairs &
$\sum_{\langle i,j\rangle} = \tfrac{1}{2}\sum_{i\neq j}$; per-f.u.\
energies carry $\tfrac12$ for like-species sums (Fe--Fe, Pt--Pt) and
no factor for Fe--Pt \\[5pt]
$J_{ij}$, $J^{\mathrm{FeFe}}_{ij}$, $J^{\mathrm{PtPt}}_{\nu\nu'}$ &
Heisenberg couplings between unit vectors & meV (mRy); to compare
with conventions in which $J$ couples moment or spin vectors, divide
by the corresponding moment magnitudes \\[5pt]
$\nu$ & site index reserved for Pt (Pd) sites & --- \\[5pt]
$\mathbf{M}_{\mathrm{Pt},\nu}$; $M^{\mathrm{FM}}_{\mathrm{Pt}}$ &
physical Pt moment; its magnitude at ferromagnetic order & $\mu_B$;
$M^{\mathrm{FM}}_{\mathrm{Pt}} \approx 0.34~\mu_B$ for FePt \\[5pt]
$\mathbf{m}_{\mathrm{Pt},\nu} =
\mathbf{M}_{\mathrm{Pt},\nu}/M^{\mathrm{FM}}_{\mathrm{Pt}}$ &
reduced Pt moment ($\mathbf{s}_\nu$ in the Polesya model) &
dimensionless; equals 1 in the FM state \\[5pt]
$J^{\mathrm{FePt}}_{i\nu}$ & Fe--Pt coupling & conjugate to the
reduced Pt moment, magnitude included [Eq.\,(\ref{eq:mry_coupling})];
coincides with the unit-vector-convention coupling in the FM state \\[5pt]
$J_{\mathrm{FePt}} \equiv \sum_i J^{\mathrm{FePt}}_{i\nu}$ &
magnitude of the FM exchange field at a Pt site & independent of
$\nu$; $1.143$~mRy $= 15.6$~meV for FePt \\[5pt]
$\mathbf{h}_\nu$; $\bar{h}(\mathbf{q})$ & exchange field on Pt site
$\nu$; its normalized magnitude & energy units;
$\bar{h}(0) = 1$ [Eqs.\,(\ref{eq:mry_coupling}),
(\ref{eq:mry_hbar})] \\[5pt]
$C$; $\lambda \equiv J_{\mathrm{FePt}}/(2C)$ & Pt on-site stiffness;
dimensionless amplitude of the Pt-mediated exchange & internally
consistent value $\lambda = 1$; bound $0 < \lambda \leq 2$
[Eq.\,(\ref{eq:mry_lambdabound})]; adopted value $3.3$ for FePt \\[5pt]
$K_{\mathrm{U}}$ & uniaxial anisotropy constant per $3d$ site &
meV/f.u.; negative = easy axis along $\mathbf{a}_3$
[Eq.\,(\ref{eq:magnetic_anisotropy})] \\[5pt]
$\mathbf{R}$; $\boldsymbol{\tau}_\mu$; $\mathbf{R}_{ij}$ & Bravais
vector; sublattice offset; bond vector connecting two sites &
Eq.\,(\ref{eq:bravais-lattice}); subscripts $i\nu$ denote Fe--Pt
bonds \\[5pt]
$\mathbf{q}$ & spiral propagation vector & quoted as a fractional
coordinate $q \in [0, 0.5]$ along $\mathbf{b}_i$ ($q = 0.5$ = zone
boundary); the physical wavevector is $q\,|\mathbf{b}_i|$ \\[5pt]
$\theta$; $\hat{\boldsymbol{\eta}}$;
$(\hat{\mathbf{u}}_1, \hat{\mathbf{u}}_2)$ & cone angle; spin-space
cone axis; transverse frame & Eq.\,(\ref{eq:spiral});
$\theta = 90^{\circ}$ (planar) unless stated \\[5pt]
$E(\mathbf{q},\theta)$; $\Delta E$ & total energy; energy relative
to the FM state & meV per formula unit (1 f.u. = 1 Fe + 1 Pt) \\[5pt]
SS; MFT (LKAG) & self-consistent spin-spiral reference;
magnetic-force-theorem couplings & Methods; LKAG = the
Green's-function realization of the MFT [Eq.\,(\ref{eqn2})] \\[5pt]
bare; dressed $\chi$ & frozen-potential response; fully
self-consistent (Dyson-resummed) response & Introduction \\[5pt]
$A$ & spin stiffness & $\mathrm{meV}\cdot\mathrm{\AA}^2$ \\[5pt]
\hline
\end{tabular}
\end{table*}

\bibliographystyle{prb-titles.bst}
\bibliography{main}

\end{document}